\documentclass[prl,nobalancelastpage,twocolumn,superscriptaddress,nofootinbib]{revtex4}

\usepackage{graphicx}
\usepackage{amsmath, amsxtra, amssymb, mathtools, nccmath, xspace}
\usepackage[utf8]{inputenc}
\usepackage{placeins} % \FloatBarrier forces to place figures
\usepackage{xcolor} %for making colored text
\usepackage{nicefrac} %this can make inline fractions
\usepackage{xspace}     % Smart spaces

\newcommand{\ket}[1]{\left|#1\right\rangle}

\newcommand{\ketbra}[2]{| #1\rangle\langle#2|}
\newcommand{\ham}{\mathcal{H}}

\newcommand{\Rb}{\ensuremath{^{87}{\rm Rb}}\xspace}
\newcommand{\tr}[1]{\textrm{tr}\left\{ #1 \right\}}
\newcommand{\mean}[1]{\langle #1 \rangle}
\newcommand{\meanmean}[1]{\langle\!\langle #1 \rangle\!\rangle}

\begin{document}
\title{
Probing many-body dynamics on a 51-atom quantum simulator	
}

\newcommand{\Caltech}{Division of Physics, Mathematics and Astronomy, California Institute of Technology, Pasadena, CA 91125, USA}
\newcommand{\Harvard}{Department of Physics, Harvard University, Cambridge, MA 02138,
USA}
\newcommand{\mitaddress}{Department of Physics and Research Laboratory of Electronics, Massachusetts Institute of Technology,
Cambridge, MA 02139, USA}
\newcommand{\Itamp}{Institute for Theoretical Atomic, Molecular and Optical Physics, Harvard-Smithsonian Center for Astrophysics, Cambridge, MA 02138, USA}

\author{Hannes Bernien}
\address{\Harvard}

\author{Sylvain Schwartz}
\address{\Harvard}
\address{\mitaddress}

\author{Alexander Keesling}
\address{\Harvard}

\author{Harry Levine}
\address{\Harvard}

\author{Ahmed Omran}
\address{\Harvard}

\author{Hannes Pichler}
\address{\Itamp}
\address{\Harvard}

\author{Soonwon Choi}
\address{\Harvard}

\author{Alexander S. Zibrov}
\address{\Harvard}

\author{Manuel Endres}
\address{\Caltech}

\author{Markus Greiner}
\address{\Harvard}

\author{Vladan Vuleti\'c}
\address{\mitaddress}

\author{Mikhail D. Lukin}
\address{\Harvard}

\begin{abstract}
    Controllable, coherent many-body systems can provide insights into the
    fundamental properties of quantum matter, enable the realization of new
    quantum phases and could ultimately lead to computational systems that
    outperform existing computers based on classical approaches. Here we
    demonstrate a method for creating controlled many-body quantum matter that
    combines deterministically prepared, reconfigurable arrays of individually
    trapped cold atoms with strong, coherent interactions enabled by excitation
    to Rydberg states. We realize a programmable Ising-type quantum spin model
    with tunable interactions and system sizes of up to 51 qubits. Within this
    model, we observe phase transitions into spatially ordered states that
    break various discrete symmetries, verify the high-fidelity preparation of
    these states and investigate the dynamics across the phase transition in
    large arrays of atoms. In particular, we observe robust manybody dynamics
    corresponding to persistent oscillations of the order after a rapid quantum
    quench that results from a sudden transition across the phase boundary. Our
    method provides a way of exploring many-body phenomena on a programmable
    quantum simulator and could enable realizations of new quantum algorithms.
\end{abstract}

\maketitle

\begin{figure}
    \includegraphics{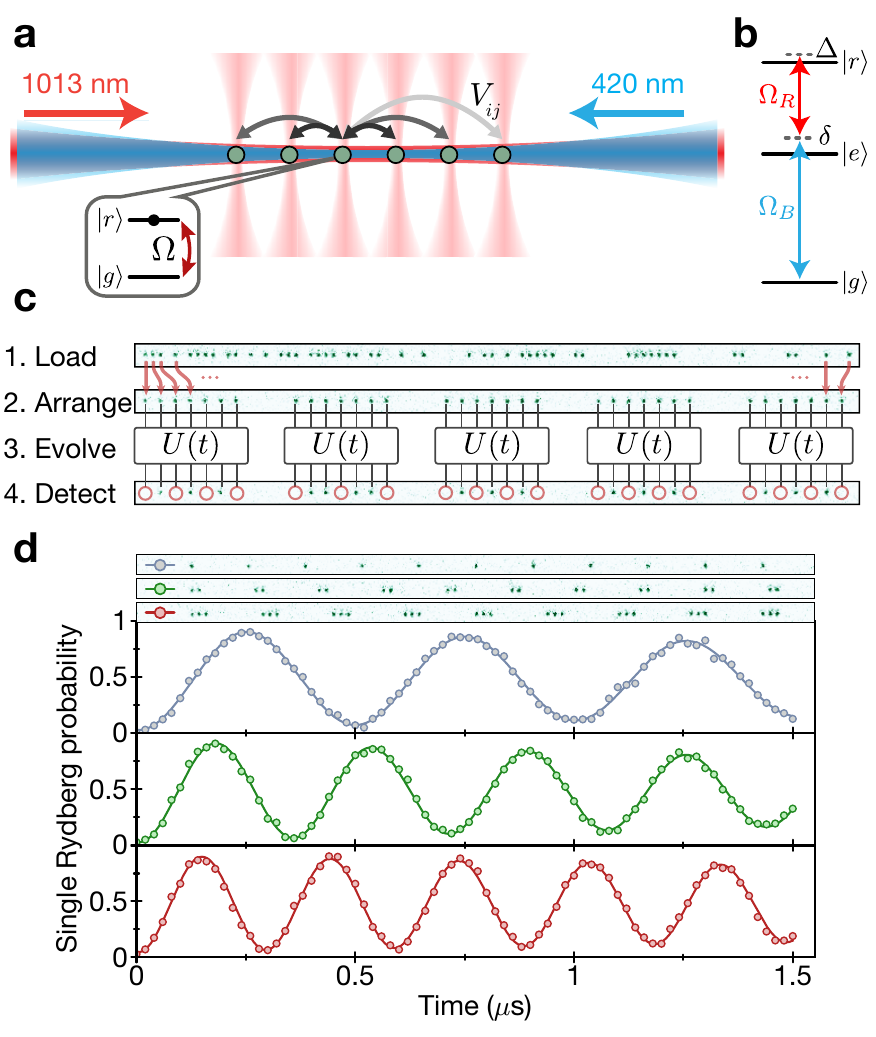}
    \caption{\textbf{Experimental platform.} \textbf{a}, Individual \Rb atoms
        are trapped using optical tweezers (vertical red beams) and arranged
        into defect-free arrays. Coherent interactions $V_{ij}$ between the
        atoms (arrows) are enabled by exciting them (horizontal blue and red
        beams) to a Rydberg state, with strength $\Omega$ and detuning $\Delta$
        (inset). \textbf{b}, A two-photon process couples the ground state
        $\ket{g}=\ket{5S_{\nicefrac{1}{2}}, F=2, m_F=-2}$ to the Rydberg state
        $\ket{r}=\ket{70S_{\nicefrac{1}{2}}, J=\nicefrac{1}{2},
        m_J=-\nicefrac{1}{2}}$ via an intermediate state
        $\ket{e}=\ket{6P_{\nicefrac{3}{2}}, F=3, m_F=-3}$ with detuning
        $\delta$, using circularly polarized 420~nm and 1013~nm lasers with
        single-photon Rabi frequencies of $\Omega_B$ and $\Omega_R$,
        respectively.  Typical experimental values are $\delta\approx 2\pi
        \times 560 \text{MHz} \gg \Omega_B, \Omega_R \approx 2\pi \times 60,
        36\,\text{MHz}$. \textbf{c}, The experimental protocol consists of
        loading the atoms into a tweezer array (1) and rearranging them into a
        preprogrammed configuration (2).  After this, the system evolves under
        $U(t)$ with tunable parameters $\Delta(t),\Omega(t)$ and $V_{ij}$. This
        evolution can be implemented in parallel on several non-interacting
        sub-systems (3).  We then detect the final state using fluorescence
        imaging (4). Atoms in state $\ket{g}$ remain trapped, whereas atoms
        in state $\ket{r}$ are ejected from the trap and detected as the absence
        of fluorescence (indicated with red circles). \textbf{d}, For resonant
        driving ($\Delta = 0$), isolated atoms (blue circles) display Rabi
        oscillations between $\ket{g}$ and $\ket{r}$.  Arranging the atoms into
        fully blockaded clusters of $N=2$ (green circles) and $N=3$ (red
        circles) atoms results in only one excitation being shared between the
        atoms in the cluster, while the Rabi frequency is enhanced by
        $\sqrt{N}$. The probability of detecting more than one excitation in the
        cluster is $\leq5\%$. Error bars indicate 68\% confidence intervals
        (CI) and are smaller than the marker size.
}
	\label{fig:experimental_platform}
\end{figure}

The realization of fully controlled, coherent many-body quantum systems is an
outstanding challenge in science and engineering. As quantum simulators,
they can provide unique insights into strongly correlated quantum systems and
the role of quantum entanglement~\cite{Bloch2012}, and enable realizations and
studies of new states of matter, even away from equilibrium. These systems also
form the basis for the realization of quantum information
processors~\cite{Ladd2010}. Although basic building blocks of such processors have
been demonstrated in systems of a few coupled
qubits~\cite{Monroe2013,Devoret2013,Awschalom2013}, the current challenge is to
increase the number of coherently coupled qubits to potentially perform tasks
that are beyond the reach of modern classical machines.

Several physical platforms are currently being explored to reach these
challenging goals. Systems composed of about 10-20 individually controlled
atomic ions have been used to create entangled states and explore quantum
simulations of Ising spin models~\cite{Monz2011,Islam2013}. Similarly sized
systems of programmable superconducting qubits have been implemented
recently~\cite{Song2017}.  Quantum simulations have been carried out in larger
ensembles of more than $100$ trapped ions without individual
readout~\cite{Gaerttner2017}. Strongly interacting quantum dynamics has been
explored using optical lattice simulators~\cite{Kuhr2016}. These systems are
already addressing computationally difficult problems in quantum
dynamics~\cite{Trotzky2012} and the fermionic Hubbard
model~\cite{Mazurenko2017}. Larger-scale Ising-like machines have been realized
in superconducting~\cite{Ronnow2014} and optical~\cite{McMahon2016} systems, but
these realizations lack either coherence or quantum nonlinearity, which are
essential for achieving full quantum speedup.

\section{Arrays of strongly interacting atoms}

A promising avenue for realizing strongly interacting quantum matter involves
coherent coupling of neutral atoms to highly excited Rydberg
states~\cite{Jaksch2000,Weimer2010_b}~(Fig.~\ref{fig:experimental_platform}a).
This results in repulsive van der Waals interactions
(of strength $V_{ij}=\nicefrac{C}{R_{ij}^6}$) between Rydberg atom pairs at a
distance $R_{ij}$~\cite{Jaksch2000}, where $C>0$ is a van der Waals
coefficient. Such interactions have recently been used to realize quantum
gates~\cite{Wilk2010,Isenhower2010,Saffman2016}, to implement strong
photon-photon interactions~\cite{Pritchard2010} and to study quantum many-body
physics of Ising spin systems in optical lattices
~\cite{Schauss2012,Schauss2015,Zeiher2017} and in probabilistically loaded
dipole trap arrays~\cite{Labuhn2016}. Our approach combines these strong,
controllable interactions with atom-by-atom assembly of arrays of cold neutral
\Rb atoms~\cite{Barredo2016,Endres2016,Kim2016}. The quantum dynamics of this
system is governed by the Hamiltonian

\begin{equation}
	\label{eq:Rydberg-Hamiltonian}
	\frac{\ham}{\hbar} = \sum_i\frac{\Omega_i}{2}\sigma^i_x-\sum_i \Delta_i n_i +\sum_{i<j}V_{ij}n_in_j,
\end{equation}
where $\Delta_i$ are the detunings of the driving lasers from the Rydberg state
(Fig.~\ref{fig:experimental_platform}b), $\sigma^i_x
=\ketbra{g_i}{r_i}+\ketbra{r_i}{g_i}$ describes the coupling between the ground
state $\ket{g_i}$ and the Rydberg state $\ket{r_i}$ of an atom at position $i$,
driven at Rabi frequency $\Omega_i$, $n_i=\ketbra{r_i}{r_i}$, and $\hbar$ is
the reduced Planck constant. Here, we focus on homogeneous coherent coupling
($\lvert\Omega_i\lvert = \Omega,\Delta_i=\Delta$), controlled by changing laser
intensities and detunings in time.  The interaction strength $V_{ij}$ is tuned
either by varying the distance between the atoms or by coupling them to a
different Rydberg state.

The experimental protocol that we implement is depicted in
Fig.~\ref{fig:experimental_platform}c (see also~\ref{fig:ED1}). First, atoms
are loaded from a magneto-optical trap into a tweezer array created by an
acousto-optic deflector. We then use a measurement and feedback procedure that
eliminates the entropy associated with the probabilistic trap loading and
results in the rapid production of defect-free arrays with more than $50$
laser-cooled atoms, as described previously~\cite{Endres2016}. These atoms are
prepared in a preprogrammed spatial configuration in a well-defined internal
ground state $\ket{g}$ (Methods). We then turn off the traps and let the system
evolve under the unitary time evolution $U(\Omega,\Delta,t)$, which is realized
by coupling the atoms to the Rydberg state $\ket{r} = \ket{70S_{1/2}}$ with
laser light along the array axis (Fig.~\ref{fig:experimental_platform}a). The
final states of individual atoms are detected by turning the traps back on, and
imaging the recaptured ground-state atoms via atomic fluorescence; the
anti-trapped Rydberg atoms are ejected. The atomic motion in the absence of
traps limits the time window for exploring coherent dynamics. For a typical
sequence duration of about $1\,\mu$s, the probability of atom loss is less than
$1$\%~(see~\ref{fig:ED2}).

The strong, coherent interactions between Rydberg atoms provide an effective
coherent constraint that prevents simultaneous excitation of nearby atoms into
Rydberg states. This  is the essence of the so-called Rydberg
blockade~\cite{Jaksch2000}, demonstrated in
Fig.~\ref{fig:experimental_platform}d.  When two atoms are sufficiently close
that their Rydberg-Rydberg interactions $V_{ij}$ exceed the effective Rabi
frequency $\Omega$, multiple Rydberg excitations are suppressed. This
defines the Rydberg blockade radius, $R_b$, at which $V_{ij} = \Omega$
($R_b=9\,\mu$m for $\ket{r}=\ket{70S_{1/2}}$ and $\Omega = 2\pi\times2\,$MHz, as
used here). In the case of resonant driving of atoms separated by $a=23\,\mu$m,
we observe Rabi oscillations associated with non-interacting atoms (blue curve
in Fig.~\ref{fig:experimental_platform}d). However, the dynamics changes
substantially as we bring multiple atoms close to each other ($a = 2.87\,\mu$m
$< R_b$). In this case, we observe Rabi oscillations between the ground state
and a collective state with exactly one excitation ($W=(1/\sqrt{N})\sum_i
\ket{g_1...r_i...g_N}$) with the characteristic $\sqrt{N}$-scaling of
the collective Rabi frequency~\cite{Dudin2012,Zeiher2015,Labuhn2016}. These
observations enable us to quantify the coherence properties of our system (see
Methods and~\ref{fig:ED3}). In particular, the amplitude of Rabi
oscillations in Fig.~\ref{fig:experimental_platform}d is limited mostly by the
state detection fidelity ($93\%$ for $\ket{r}$ and $\sim98\%$ for $\ket{g}$;
Methods). The individual Rabi frequencies are controlled to better than $3\%$
across the array, whereas the coherence time is limited ultimately by the small
probability of spontaneous emission from the intermediate state $\ket{e}$
during the laser pulse (scattering rate $0.022/\mu$s; Methods).

 \section{Programmable quantum simulator}

\begin{figure*}
	\includegraphics{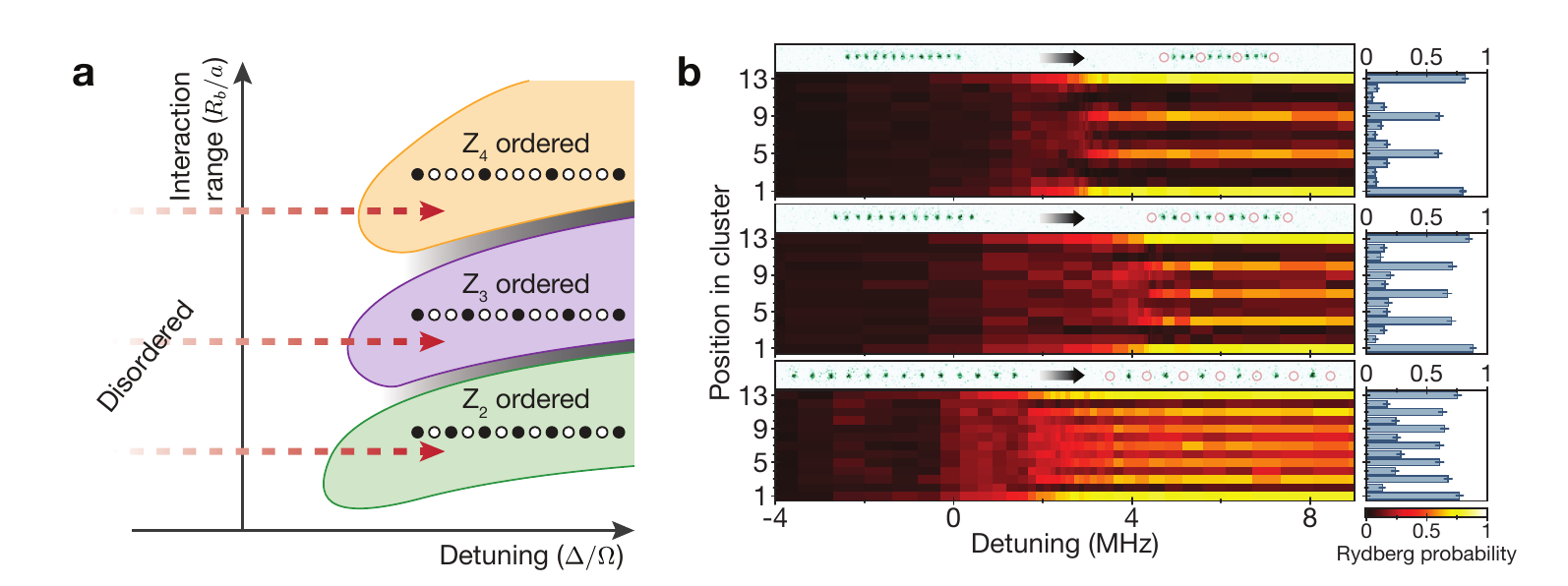}
	\caption{
    \textbf{Phase diagram and build-up of crystalline phases.} \textbf{a}, A
schematic of the ground-state phase diagram of the
Hamiltonian in equation~(\ref{eq:Rydberg-Hamiltonian}) displays phases with
various broken symmetries depending on the interaction range $R_b/a$ ($R_b$,
blockade radius; $a$, trap spacing) and detuning $\Delta$ (see main text).
Shaded areas indicate potential incommensurate phases (diagram adapted
from~\cite{Fendley2004}). Here we show the experimentally accessible region;
further details can be found
in~\cite{Fendley2004,Sachdev2002,Schachenmayer2010}.  \textbf{b},The build-up of
Rydberg crystals on a 13-atom array is observed by slowly changing the laser
parameters, as indicated by the red dashed arrows in \textbf{a} (see also
Fig.~\ref{fig:preparation_fidelity}a). The bottom panel shows a configuration
in which the atoms are $a=5.74\,\mu$m apart, which results in a nearest neighbour
interaction of $V_{i,i+1}=2\pi\times24\,$MHz and leads to a Z$_2$ order whereby
every other atom is excited to the Rydberg state $\ket{r}$. The bar plot
on the right displays the final, position-dependent Rydberg probability (error
bars denote 68\% confidence intervals). The configuration in the middle panel ($a = 3.57\,\mu$m,
$V_{i,i+1}=2\pi\times414.3\,$MHz) results in Z$_3$ order and the top panel
($a=2.87\,\mu$m, $V_{i,i+1}=2\pi\times1536\,$MHz) in Z$_4$ order. For
each configuration, we show a single-shot fluorescence image before (left) and
after (right) the pulse. Red circles highlight missing atoms, which are
attributed to Rydberg excitations.}
	\label{fig:different_crystal_orders}
\end{figure*}

In the case of homogeneous coherent coupling considered here, the Hamiltonian
in equation~(\ref{eq:Rydberg-Hamiltonian}) resembles closely the paradigmatic
Ising model for effective spin-1/2 particles with variable interaction range.
Its ground state exhibits a rich variety of many-body phases that break
distinct spatial symmetries (Fig.~\ref{fig:different_crystal_orders}a).
Specifically, at large negative values of $\nicefrac{\Delta}{\Omega}$, its
ground state corresponds to all atoms in the state $\ket{g}$, corresponding to
a paramagnetic or disordered phase. As $\nicefrac{\Delta}{\Omega}$ is increased
towards large positive values, the number of atoms in $\ket{r}$ increases and
interactions between them become important. This gives rise to spatially
ordered phases in which Rydberg atoms are arranged regularly across the array,
resulting in `Rydberg crystals' with different spatial
symmetries~\cite{Fendley2004,Pohl2010}, as illustrated in
Fig.~\ref{fig:different_crystal_orders}a. The origin of these correlated states
can be understood intuitively by first considering the situation in which
$V_{i,i+1} \gg \Delta \gg \Omega \gg V_{i,i+2}$, that is, with blockade for
neighbouring atoms but negligible interaction between next-nearest neighbours. In
this case, the ground state corresponds to a Rydberg crystal that breaks Z$_2$
translational symmetry in a manner analogous to antiferromagnetic order in
magnetic systems.  Moreover, by tuning the parameters such that $V_{i,i+1},
V_{i,i+2} \gg \Delta \gg \Omega \gg V_{i,i+3}$ and $V_{i,i+1}, V_{i,i+2},
V_{i,i+3} \gg \Delta \gg \Omega \gg V_{i,i+4}$, we obtain arrays with broken
Z$_3$ and Z$_4$ symmetries,
respectively~(Fig.~\ref{fig:different_crystal_orders}).

To prepare the system in these phases, we control the detuning $\Delta(t)$ of
the driving lasers dynamically to adiabatically transform the ground state of
the Hamiltonian from a product state of all atoms in $\ket{g}$ to crystalline
states~\cite{Pohl2010,Schauss2015}. In contrast to prior work where Rydberg
crystals are prepared via a sequence of avoided
crossings~\cite{Pohl2010,Schauss2015,Petrosyan2016}, the operation at a finite
$\Omega$ and well-defined atom separation allows us to move across a
single phase transition into the desired phase directly~\cite{Schachenmayer2010}.

\begin{figure}
	\includegraphics{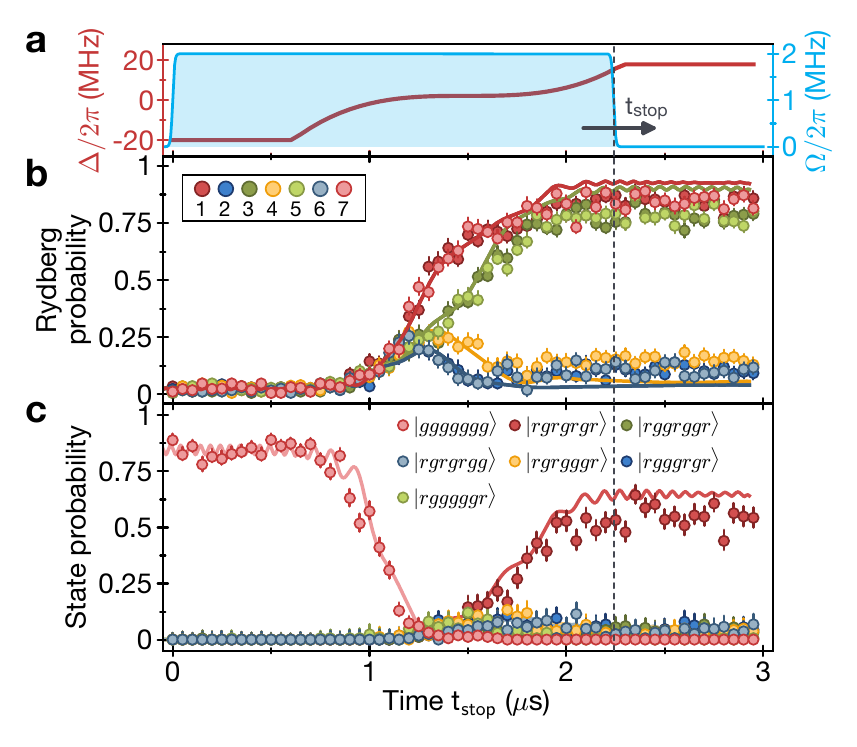}
	
    \caption{\textbf{Comparison with a fully coherent simulation.} \textbf{a},
        The laser driving consists of a square shaped pulse $\Omega(t)$ (blue)
        with a detuning $\Delta(t)$ (red) that is chirped from negative to
        positive values. \textbf{b}, The data show the time evolution of the
        Rydberg excitation probability for each atom in a 7-atom cluster
        (colored points), obtained by varying the stopping time $t_{\rm stop}$ of laser excitation
        the laser-excitation pulse $\Omega(t)$. The corresponding curves are
        theoretical single atom-trajectories obtained from an exact simulation of
        quantum dynamics with equation~(\ref{eq:Rydberg-Hamiltonian}), the functional
        form of $\Delta(t)$ and $\Omega(t)$ used in the experiment, and finite
        detection fidelity.  \textbf{c}, Evolution of the seven most probable
        many-body states (data). The target state is reached with $54(4)$\%
        probability ($77(6)$\% when corrected for finite detection fidelity). Solid
        lines are theoretical (simulated) many-body trajectories. Error
        bars in b and c denote 68\% confidence intervals.}
\label{fig:preparation_fidelity}
\end{figure}

In the experiment, we first prepare all atoms in state $\ket{g}
=\ket{5S_{\nicefrac{1}{2}}, F=2, m_F = -2}$ by optical pumping. We then switch
on the laser fields and sweep the two-photon detuning from negative to positive
values using a functional form shown in Fig.~\ref{fig:preparation_fidelity}a.
Fig.~\ref{fig:different_crystal_orders}b displays the resulting single atom
trajectories in a group of 13~atoms for three different interaction strengths
as we vary the detuning $\Delta$. In each of these instances, we observe a
clear transition from the initial state $\ket{g_1,...,g_{13}}$ to an ordered
state of different broken symmetry. The distance between the atoms determines
the interaction strength which, leads to different crystalline orders for a
given final detuning. To achieve a Z$_2$ order, we arrange the atoms with a
spacing of $5.74\,\mu$m, which results in a measured nearest-neighbour
interaction (see~\ref{fig:ED4}) of $V_{i,i+1} =
2\pi\times24\,$MHz~$\gg\Omega=2\pi\times2\,$MHz, while the next-nearest
neighbour interaction is small ($2\pi\times0.38\,$MHz). This  results in a
build-up of antiferromagnetic order whereby every other trap site is occupied by a
Rydberg atom (Z$_2$ order). By reducing the spacing between the atoms to
$3.57\,\mu$m and $2.87\,\mu$m,  Z$_3$- and Z$_4$- order is observed,
respectively (Fig.~\ref{fig:different_crystal_orders}b). 
\\

We benchmark the performance of the quantum simulator by comparing the measured
build-up of Z$_2$ order with theoretical predictions for a $N=7$ atom system,
obtained via exact numerical simulations. As shown in
Fig.~\ref{fig:preparation_fidelity}, this fully coherent simulation without
free parameters yields excellent agreement with the observed data when the
finite detection fidelity is accounted for. The evolution of the many-body
states in Fig.~\ref{fig:preparation_fidelity}c shows that we measure the
perfect antiferromagnetic state with $54(4)$\% probability (here and elsewhere,
unless otherwise specified, the error denotes the 68\% confidence interval).
When corrected for the known detection infidelity, we find that the desired
many-body state is reached with probability $P=77(6)$\%. 

\begin{figure}
	\includegraphics{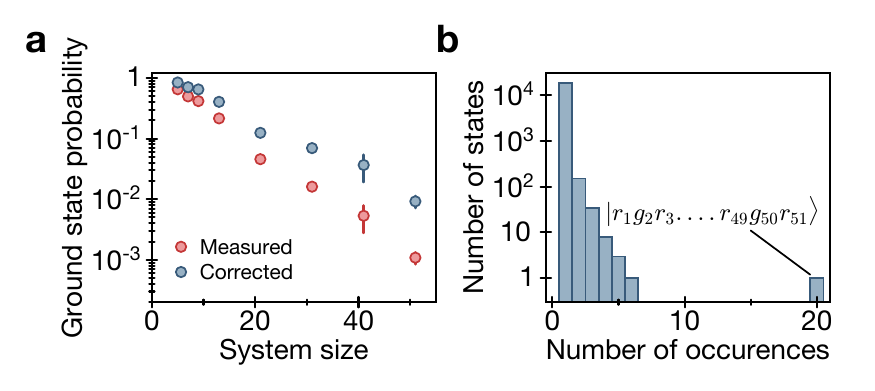}
	
    \caption{\textbf{Scaling behavior.} \textbf{a}, Preparation fidelity of the
        crystalline ground state as a function of cluster size. The red circles
        are the measured values and the blue circles are corrected for finite
        detection fidelity (Methods). Error bars denote 68\% confidence
        intervals.  \textbf{b}, Number of observed many-body states per number of
        occurrences out of 18439 experimental realizations in a 51-atom cluster. The
        most frequently occurring state $\ket{r_1 g_2 r_3 \cdots r_{49} g_{50} r_{51}}$ is the ground state of the
        many-body Hamiltonian.}
	\label{fig:size-scaling}
\end{figure}

To investigate the way in which the preparation fidelity depends on system
size, we perform detuning sweeps on arrays of various sizes
(Fig.~\ref{fig:size-scaling}a). We find that the probability of observing the
system in the many-body ground state at the end of the sweep decreases as the
system size is increased. However, even at system sizes as large as 51~atoms,
the perfectly ordered crystalline many-body state is obtained with
$P=0.11(2)\%$  ($P=0.9(2)\%$ when corrected for detection fidelity). These
probabilities compare favorably with those measured previously for smaller
systems~\cite{Islam2013,Richerme2013}~(see also~\ref{fig:ED5}) and are
remarkably large in view of the exponentially large, $2^{51}$-dimensional
Hilbert space of the system. Furthermore, we find that the state with perfect
Z$_2$ order is by far the most commonly observed many-body state
(Fig.~\ref{fig:size-scaling}b).  The observations of perfectly ordered states
resulting from the dynamical evolution across the phase transition indicate
that a substantial degree of quantum coherence is preserved in our 51 atom
system over the entire evolution time.

\section{Quantum dynamics across a phase transition}
 
\begin{figure*}
	\includegraphics{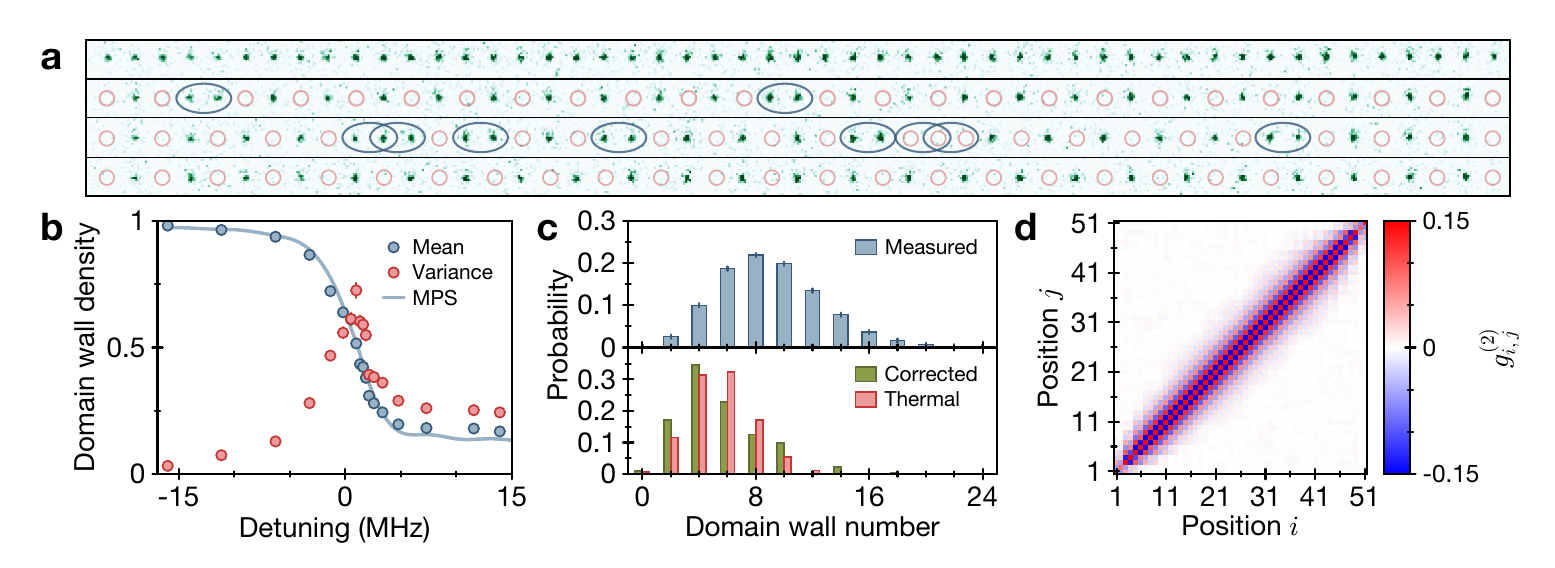}
    \caption{\textbf{Quantifying Z$_2$ order in a 51-atom array after a slow
        detuning sweep.} \textbf{a}, Single-shot fluorescence images of a
        51-atom array before applying the adiabatic pulse (top row) and after
        the pulse (bottom three rows correspond to three separate instances).
        Red circles mark missing atoms, which are attributed to Rydberg
        excitations. Domain walls are identified as either two neighbouring
        atoms in the same state or a ground state atom at the edge of the array
        (Methods), and are indicated with blue ellipses. Long Z$_2$ ordered chains
        between domain walls are observed. \textbf{b}, Blue points show the
        mean of the domain-wall density as a function of detuning during the
        sweep. Error bars show the standard error of the mean and are smaller
        than the marker size. The red circles are the corresponding variances,
        and the error bars represent one standard deviation. The onset of the
        phase transition is indicated by a decrease in the domain-wall density
        and a peak in the variance (see main text for details). Each point is
        obtained from about $1000$ realizations. The solid blue curve is a fully
        coherent matrix product state (MPS) simulation without free parameters (bond dimension
        $D=256$), taking measurement fidelities into account. \textbf{c},
        Domain wall number distribution for $\Delta = 2\pi\times14\,$MHz,
        obtained from 18439 experimental realizations (blue bars, top).
        Error bars indicate 68\% confidence intervals. Owing to the boundary
        conditions, only even numbers of domain walls can appear (Methods).
        Green bars (bottom) show the distribution obtained by
        correcting for finite detection fidelity using a maximum-likelihood
        method (Methods), which results in an average number of $5.4$~domain
        walls; red bars show the distribution of a thermal state  with the same
        mean domain wall density (Methods). \textbf{d}, Measured correlation
        function~(\ref{eq:corr_function}) in the Z$_2$ phase.
	}
	\label{fig:large_system_size}
\end{figure*} 

We next present a detailed study of the transition into the Z$_2$ phase in an
array of 51~atoms, which allows us to minimize edge effects and study the
properties of the bulk. We first focus on analysing the atomic states that result
from a slow sweep of the laser detuning across the resonance, as described in
the previous section~(Fig.~\ref{fig:large_system_size}). In single instances of
the experiment, after such a slowly changing laser pulse, we observe long
ordered chains where the atomic states alternate between Rydberg and ground
state. These ordered domains can be separated by domain walls that consist of
two neighbouring atoms in the same electronic state
(Fig.~\ref{fig:large_system_size}a)~\cite{Sachdev2009}. These
features cannot be observed in the average excitation probability of the
bulk~(\ref{fig:ED6}a). 

The domain-wall density can be used to quantify the transition from the
disordered phase into the ordered Z$_2$ phase as a function of detuning
$\Delta$. As the system enters the Z$_2$ phase, ordered domains grow in size,
leading to a substantial reduction in the domain wall density (blue points in
Fig.~\ref{fig:large_system_size}b).  Consistent with expectations for an
Ising-type second-order quantum phase transition~\cite{Sachdev2009}, we observe
domains of fluctuating lengths close to the transition point between the two
phases, which is reflected by a pronounced peak in the variance of the
domain-wall density.  Consistent with predictions from finite-size scaling
analysis~\cite{Fendley2004,Sachdev2002}, this peak is shifted towards positive
values of $\nicefrac{\Delta}{\Omega}$. The measured position of the peak is
$\Delta\approx0.5\Omega$.  The observed domain-wall density is in excellent
agreement with fully coherent simulations of the quantum dynamics based on
51-atom matrix product states (blue line in Fig.~\ref{fig:large_system_size}b);
however, these simulations underestimate the variance at the phase transition
(see~\ref{fig:ED6}b).

At the end of the sweep, deep in the Z$_2$ phase
($\nicefrac{\Delta}{\Omega}\gg1$) we can neglect $\Omega$ so that the
Hamiltonian in equation~(\ref{eq:Rydberg-Hamiltonian}) becomes essentially classical. In
this regime, the measured domain wall number distribution enables us to
infer directly the statistics of excitations created when crossing the phase transition.
From 18439 experimental realizations we obtain the distribution depicted in
Fig.~\ref{fig:large_system_size}c with an average of $9.01(2)$ domain walls.
From a maximum-likelihood estimation we obtain the distribution corrected for
detection fidelity (see~\ref{fig:ED7}), which corresponds to a state that has
on average $5.4$~domain walls.  These domain walls are probably created as a result
of non-adiabatic transitions from the ground state when crossing the phase
transition~\cite{Zurek2005}, where the energy gap depends on the system size
(and scales as $1/N$)~\cite{Sachdev2002}.  In addition, the preparation
fidelity is limited by spontaneous emission during the laser pulse (an
average number of 1.1 photons is scattered per $\mu$s for the entire array;
see Methods).

To further characterize the created Z$_2$ ordered state, we evaluate the
correlation function
\begin{equation}
    \label{eq:corr_function}
g^{(2)}_{ij}=\mean{ n_i n_{j} } - \mean{n_i} \mean{n_{j}}
\end{equation}
where the average $\mean{\cdots}$ is taken over experimental repetitions. We
find that the correlations decay exponentially over distance with a decay
length of $\xi= 3.03(6)$~sites (see Fig.~\ref{fig:large_system_size}d and
Methods; the error denotes the uncertainty in the fit). We note that this
length does not fully characterize the system as discussed below (see
also~\ref{fig:ED8}).

\begin{figure*}
	\includegraphics{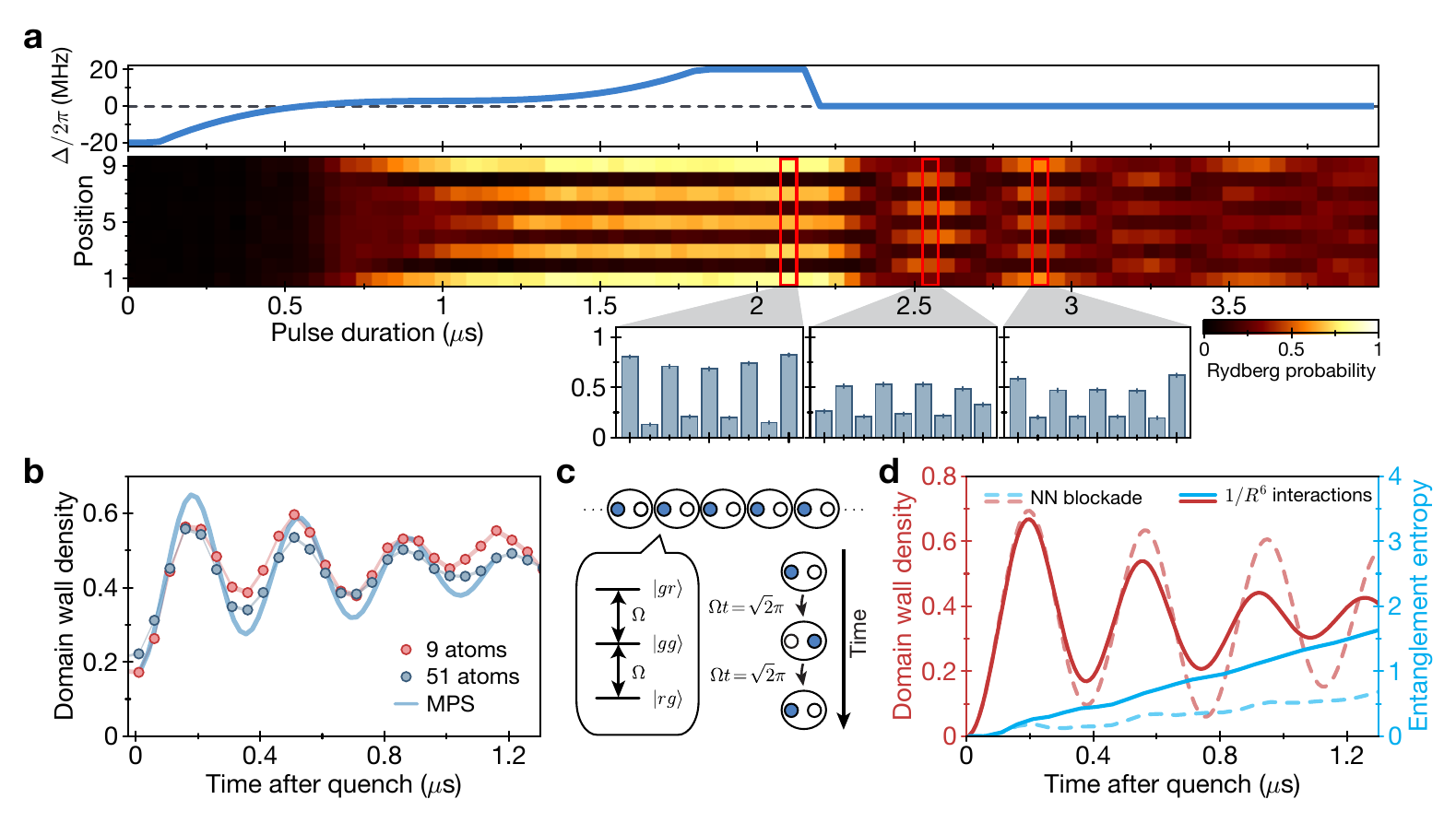}
	
    \caption{\textbf{Emergent oscillations in many-body dynamics after sudden
    quench.} \textbf{a}, Schematic sequence (top, showing  $\Delta(t)$)
    involves adiabatic preparation and then a sudden quench to single-atom
    resonance. The single-atom trajectories are shown (bottom) for a 9 atom
    cluster, with the colour scale indicating the Rydberg probability. We
    observe that the initial crystal with a Rydberg excitation at every odd
    trap site (left inset) collapses after the quench, and a crystal with
    an excitation at every even site builds up (middle inset). At a later time the
    initial crystal revives with a frequency of $\Omega/1.38(1)\,$ (right inset).
    Error bars denote 68\% confidence intervals. \textbf{b}, Domain-wall density after the
    quench. The dynamics decay slowly on a timescale of 0.88 $\mu$s. Shaded
    region represents the standard error of the mean. Solid blue line is a fully
    coherent matrix product state (MPS) simulation with bond dimension $D=256$,
    taking into account measurement fidelity. \textbf{c}, Toy model of
    non-interacting dimers (see main text). Blue (white) circles represent atoms in
    state $\ket{g}$ ($\ket{r}$). \textbf{d}, Numerical calculations
    of the dynamics after a quench, starting from an ideal 25-atom crystal,
    obtained from exact diagonalization. Domain-wall density (red) 
    and the growth of entanglement entropy of the half chain (13 atoms; blue)
    are shown as functions of time after the quench. Dashed lines take into
    account only nearest-neighbour (NN) blockade constraint. Solid lines correspond
    to the full 1$/R^6$ interaction potential.}
	\label{fig:crystal_dynamics}
\end{figure*}

Finally, Fig.~\ref{fig:crystal_dynamics} demonstrates that our approach  also
enables the study of coherent dynamics of many-body systems far from
equilibrium. Specifically, we focus on the quench dynamics of Rydberg crystals
initially prepared deep in the Z$_2$ ordered phase, as we change the
detuning $\Delta(t)$ suddenly to the single-atom resonance $\Delta =0$
(Fig.~\ref{fig:crystal_dynamics}a).  After such a quench, we observe
oscillations of many-body states between the initial crystal and a
complementary crystal in which each internal atomic state is inverted
(Fig.~\ref{fig:crystal_dynamics}a). Remarkably, we find that these oscillations are
robust, persisting over several periods with a frequency that is
largely independent of the system size. This is confirmed by measuring the
dynamics of the domain wall density, which signals the appearance and disappearance
of the crystalline states, shown in Fig.~\ref{fig:crystal_dynamics}b for arrays
of 9 and 51~atoms. We find that the initial crystal repeatedly revives with a
period that is slower by a factor of 1.38(1) (error denotes the uncertainty in the fit)
compared to the Rabi-oscillation period for independent, non-interacting atoms. 

\section{Discussion} 

Several important features of these experimental observations should be noted.
First, the Z$_2$ ordered state cannot be characterized by a simple
thermal ensemble. More specifically, if an effective temperature is estimated
based on the experimentally determined, corrected domain wall density of~0.1,
then the corresponding thermal ensemble predicts a correlation length $\xi_{\rm th}
= 4.48(3)$, which is significantly longer than the measured value $\xi=
3.03(6)$ (Methods). Such a discrepancy is also reflected in distinct
probability distributions for the number of domain
walls~(Fig.~\ref{fig:large_system_size}c).  These observations suggest that the
system does not thermalize within the timescale of the Z$_2$ state preparation.

Even more striking is the coherent and persistent oscillation of the
crystalline order after the quantum quench. With respect to the quenched
Hamiltonian ($\Delta = 0$), the energy density of our Z$_2$ ordered state
corresponds to that of an infinite-temperature ensemble within the manifold
constrained by Rydberg blockade. Also, our Hamiltonian does not have any
explicit conserved quantities other than total energy. Nevertheless, the
oscillations persist well beyond the natural timescale of local relaxation
($1/\Omega$) as well as the fastest timescale, $1/V_{i,i+1}$.

To understand these observations, we consider a simplified model in which the
effect of long-range interactions is neglected, and nearest-neighbour
interactions are replaced by hard constraints on neighbouring excitations of
Rydberg states~\cite{Fendley2004}. In this limit, the qualitative behavior of
the quench dynamics can be understood in terms of dimerized spins
(Fig.~\ref{fig:crystal_dynamics}c); owing to the blockade constraint, each
dimer forms an effective spin-1 system with three states ($\ket{rg}$,
$\ket{gg}$ and $\ket{gr}$), in which the resonant drive ``rotates'' the three
states over the period $\sqrt{2} (2\pi/\Omega)$, close to that observed
experimentally. Although this qualitative picture does not take into account the
strong interactions (constraints) between neighbouring dimers, it  can be
extended by considering a minimal variational ansatz for the many-body wave
function based on matrix product states that respect all blockade constraints
(Methods). Using the time-dependent variational principle, we derive analytical
equations of motion and obtain a crystalline-order oscillation with a frequency
of about $\Omega/1.51$ (see~\ref{fig:ED9}), which is within 10\% of the
experimental observations. These considerations are  supported by various
numerical simulations. The exact numerics predict that this simplified
model exhibits crystal oscillations with the observed frequency, while the
entanglement entropy grows at a rate much smaller than $\Omega$, indicating
that the oscillation persists over many cycles
(Fig.~\ref{fig:crystal_dynamics}d and Methods). The addition of long-range
interactions leads to a faster decay of the oscillations, with a timescale that
is determined by $1/V_{i,i+2}$, in good agreement with experimental
observations~(Fig.~\ref{fig:crystal_dynamics}b); the entanglement entropy
also grows on this timescale~(Fig.~\ref{fig:crystal_dynamics}d, see
also~\ref{fig:ED10}).

Our observations and analysis indicate that the decay of crystal oscillation is
governed by weak next-nearest-neighbour interactions. This relatively slow
thermalization is rather unexpected, because our Hamiltonian, with or without
long-range interactions, is far from any known integrable
system~\cite{Fendley2004}, and features neither strong disorder nor explicitly
conserved quantities~\cite{DAlessio2016}. Instead, our observations  are
probably associated with constrained dynamics due to Rydberg blockade and large
separations of timescales ($V_{i,i+1} \gg \Omega \gg
V_{i,i+2}$~\cite{Abanin2016}) that result in an effective Hilbert-space
dimension that is determined by the golden ratio
$(1+\sqrt{5})^N/2^N$~\cite{Feiguin2007,Lesanovsky2012}. The evolution within
such a constrained Hilbert space gives rise to the so-called quantum dimer
models, which are known to possess non-trivial dynamics~\cite{Moessner2011}.
Although these considerations provide important insights into the origin of
robust emergent dynamics, our results challenge conventional theoretical
concepts and so warrant further studies.

\section{Outlook}
Our observations demonstrate that Rydberg excitation of arrays of neutral atoms
is a promising way of studying quantum dynamics and quantum simulations in
large systems. Our method can be extended and improved in several ways.
Individual qubit rotations around the $z$ axis can be implemented using light
shifts associated with trap light, while a second acousto-optic deflector could be used for
individual control of coherent rotations around other directions. Further
improvement in coherence and controllability could be obtained by encoding qubits
into hyperfine sublevels of the electronic ground state and using
state-selective Rydberg excitation~\cite{Zeiher2017}.  Implementing
two-dimensional arrays could provide a path towards realizing thousands of
traps. Such two-dimensional configurations could be realized by using
a two-dimensional acousto-optic deflector directly or by creating a static two-dimensional
lattice of traps and sorting atoms with an independent acousto-optic deflector, as demonstrated
recently~\cite{Barredo2016}. With increased
loading efficiencies~\cite{Lester2015}, the robust creation and control of
arrays of hundreds of atoms is feasible. 

Although our current observations already provide insights into
the physics associated with transitions into ordered phases and enable us to explore
new many-body phenomena in quantum dynamics, they can be extended
along several directions~\cite{Weimer2010_b}. These include studies of various
aspects of many-body coherence and entanglement in large
arrays~\cite{Pichler2016}, investigation of quantum critical dynamics and tests
of the quantum Kibble-Zurek hypothesis~\cite{Zurek2005}, and the exploration of
stable non-equilibrium phases of matter~\cite{Schiulaz2015}. Further extension
may allow for studies of the interplay between long-range interactions and
disorder, of quantum scrambling~\cite{Swingle2016}, of topological states in spin
systems~\cite{Chandran2016}, of the dynamics of Fibonacci
anyons~\cite{Feiguin2007,Lesanovsky2012}, and of chiral clock
models associated with transitions into exotic Z$_3$ and Z$_4$
states~\cite{Huse1984}. Finally, we note that our approach is well suited for
the realization and testing of quantum optimization
algorithms~\cite{Lechner2015,Farhi2016} with system sizes that cannot be
simulated by modern classical machines.

\section{Acknowledgements}
We thank E. Demler, A. Chandran, S. Sachdev, A. Vishwanath, P. Zoller, P. Silvi, T. Pohl, M. Knap, M. Fleischhauer, S. Hofferberth and A. Harrow for insightful discussions. This work was supported by NSF, CUA, ARO, and Vannevar Bush Faculty Fellowship. H.B. acknowledges support by a Rubicon Grant of the Netherlands Organization for Scientific Research (NWO). A.O. acknowledges support by a research fellowship from the German Research Foundation (DFG). S.S. acknowledges funding from the European Union under the Marie Sk\l odowska Curie Individual Fellowship Programme H2020-MSCA-IF-2014 (project number 658253). H.P. acknowledges support by the National Science Foundation (NSF) through a grant at the Institute of Theoretical Atomic Molecular and Optical Physics (ITAMP) at Harvard University and the Smithsonian Astrophysical Observatory. H.L. acknowledges support by the National Defense Science and Engineering Graduate (NDSEG) Fellowship.

\section{Author Contributions} The experiments and data analysis were carried out by H.B., S.S., A.K., H.L., A.O., A.S.Z., and M.E.. Theoretical analysis was performed by H.P. and S.C.. All work was supervised by M.G., V.V. and  M.D.L.. All authors discussed the results and contributed to the manuscript.

\section{Author Information} Reprints and permissions information is available at www.nature.com/reprints. The authors declare no competing financial interests. Readers are welcome to comment on the online version of the paper. Correspondence and requests for materials should be addressed to M.L. (lukin@physics.harvard.edu).

\FloatBarrier

\bibliographystyle{naturemag}

{
\renewcommand{\addcontentsline}[3]{}

}

%---------------------Supplementary:

\clearpage
\newpage

{
\renewcommand{\addcontentsline}[3]{}
\section{Methods}
}

\setcounter{figure}{0}
\makeatletter 
\renewcommand{\thefigure}{Extended Data Fig. \@arabic\c@figure}
\renewcommand{\fnum@figure}{\textbf{Extended Data Figure \@arabic\c@figure}}
\makeatother

\begin{footnotesize}

\noindent
\textbf{Trapping set-up and experimental sequence.}
Our set-up consists of a linear array of up to 101 evenly spaced optical
tweezers. The tweezers are generated by feeding a multi-tone radio-frequency signal into an
acousto-optic deflector (AA Opto-Electronic model DTSX-400-800.850), generating
multiple deflections in the first diffraction order and focusing them into the
vacuum chamber using a 0.5 NA objective (Mitutoyo G Plan Apo 50X). The beams
have a wavelength of $808\,$nm and a waist of approximately $0.9\,\mu$m,
forming traps of approximate depth $1\,$mK.

A diagram of the experimental sequence is shown in~\ref{fig:ED1}a. The traps
are loaded from a magneto-optical trap, leading to individual tweezer
single-atom loading probabilities of around $0.6$. A fluorescence image of the
array is taken, and the empty traps are turned off; the filled traps are
rearranged to bring the atoms into their preprogrammed
positions~\cite{Endres2016}. After the rearrangement procedure, another
image of the array is taken to preselect on instances in which the initial
configuration is defect-free. After taking the second image, we apply a
magnetic field of about $1.5\,$G along the axis of the array and then
optically pump all atoms into the $\ket{F=2, m_F = -2}$ state using a
$\sigma^{-}$-polarized beam resonant to the
$\ket{5S_{1/2},F=2}\to\ket{5P_{3/2},F=2}$ transition. We then turn off the
traps, pulse the Rydberg lasers on a timescale of a few microseconds, then
turn the traps back on to recapture the atoms that are in the ground state
$\ket{g}$ while pushing away the atoms in the Rydberg state $\ket{r}$, and
finally take a third image. Because of their long lifetime, most of the Rydberg
atoms escape from the trapping region before they decay back to the ground
state. This provides a convenient way to detect them as missing atoms on the
third image (with finite detection fidelity discussed in Methods section `State
detection fidelity'). The
entire experimental sequence, from magneto-optical trap formation to the third
image, takes approximately $250\,$ms.

\noindent
\textbf{Rydberg laser set-up.}
To introduce interactions within the array, we couple the atomic ground state
$\ket{g}=\ket{5S_{1/2}, F=2, m_F = -2}$ to a target Rydberg state
$\ket{r}=\ket{70S_{1/2}, m_J = -1/2}$. The van der Waals interaction between
two \Rb $70S$ atoms follows a $1/R^6$ power law and is on the order of $1\,$MHz
at $10\,\mu$m~\cite{Singer2005}, making it the dominant energy scale in our
system for up to several lattice sites.

The coupling between states $\ket{g}$ and $\ket{r}$ is induced by a two-photon
transition, with $\ket{6P_{3/2}}$ as the intermediate level. We drive the
transition between $\ket{g}$ and $\ket{6P_{3/2}}$ with a blue $420\,$nm laser
(MOGLabs cateye diode laser CEL002) and the transition between $\ket{6P_{3/2}}$
and $\ket{r}$ with an infrared $1013\,$nm laser injecting a tapered amplifier
(MOGLabs CEL002 and MOA002). The detuning $\delta$ of the blue laser from the
$\ket{g} \leftrightarrow \ket{6P_{3/2}}$ transition is chosen to be much larger
than the single-photon Rabi frequencies (typically $\delta\approx 2\pi \times
560\,$MHz $\gg (\Omega_B, \Omega_R) \approx 2\pi \times (60, 36)\,$MHz, where
$\Omega_{B}$ and $\Omega_{R}$ are the single-photon Rabi frequencies for the
blue and red lasers, respectively), such that the dynamics can be safely
reduced to a two-level transition $\ket{g} \leftrightarrow \ket{r}$ driven by
an effective Rabi frequency $\Omega=\Omega_{B} \Omega_{R}/(2\delta)\approx 2\pi
\times 2\,$MHz. 

The blue and IR beams are applied counter-propagating to one another along the
axis of the array. An external magnetic field is applied in addition, and the
beams are circularly polarized such that blue laser drives the $\sigma^{-}$
transition between $\ket{g}$ and $\ket{e}=\ket{6P_{3/2}, F=3, m_F = -3}$, while
the red laser drives the $\sigma^{+}$ transition between $\ket{e}$ and
$\ket{r}$. Such a stretched configuration minimizes the probability of exciting
unwanted states such as $\ket{70S_{1/2}, m_J = +1/2}$. The two beams are
focused to waists of $20\,\mu$m (blue) and $30\,\mu$m (infrared) at the position of
the atoms, to get high intensity while still being able to
couple all atoms in the array homogeneously (see Methods section `Coherence
limitations').

The Rydberg lasers interact with the atoms during one experimental cycle for a
few microseconds. To maintain laser coherence, the line width must be
much smaller than a few tens of kilohertz. To achieve this, we use a fast
Pound-Drever-Hall scheme to lock our Rydberg lasers to an ultralow-expansion
reference cavity (ATF-6010-4 from Stable Laser Systems, with a finesse of
$\geq4000$ at both $420\,$nm and $1013\,$nm). The optical set-up used for this
purpose is sketched in~\ref{fig:ED1}b. A fraction of the beam from the blue
laser first goes through a phase modulator (Newport 4005) driven by a $18\,$MHz
sinusoidal signal, before being coupled to a longitudinal mode of the reference
cavity. The reflected beam from the cavity is sent to a fast photodetector
(Thorlabs PDA8A), whose signal is demodulated and low-pass-filtered to create
an error signal which is fed into a high-bandwidth servo controller (Vescent
D2-125). The feedback signal from the servo controller is applied to the
current of the laser diode using a dedicated fast-input port on the laser
headboard. The measured overall bandwidth of the lock is on the order of
$1\,$MHz. The other part of the blue laser beam goes through an acousto-optic
modulator (IntraAction ATM-1002DA23), whose first diffraction order is used to
excite atoms, providing frequency and amplitude control for the Rydberg pulses.

A similar scheme is implemented for the $1013\,$nm laser, with a notable
difference that the beam used for the frequency lock first goes through a
high-bandwidth ($>5\,$GHz) fiber-based electro-optic modulator (EOSpace
PM-0S5-05-PFA-PFA-1010/1030). Rather than the carrier, we use a first-order
sideband from the electro-optic modulator for the lock, which makes it possible
to tune the frequency of the red laser over a full free-spectral range of the
reference cavity ($1.5\,$GHz) by tuning the driving frequency of the
high-bandwidth electro-optic modulator. Following~\cite{Hall1992}
and~\cite{Fox2003}, we estimate that the contribution to the line width of the
laser of the noise within the servo loop relative to the cavity is less than
$500\,$Hz.

\noindent
\textbf{Measuring interaction strengths.} We measure experimentally the $70S
\rightarrow 70S$ van der Waals interactions between atom pairs separated by
$5.74~\mu$m (identical to the spacing used for observing the Z$_2$ ordered
phase) to confirm our estimate of interaction strengths and to provide
independent (and more precise) estimation of the exact atom
spacing~(\ref{fig:ED4}). At this spacing we expect the interaction $V$ to be 
about $2\pi \times 20$ MHz. We apply our two laser fields (420 nm and
1013 nm) to couple each atom to the Rydberg state, with two-photon detuning
$\Delta$. For $\Delta=0$, we observe resonant coupling from $\ket{g,g}$ to
$\ket{W} = (\ket{g,r} + \ket{r,g})/\sqrt{2}$, as expected for the
blockaded regime in which $\Omega = 2\pi \times 2$ MHz  $\ll V$. However, there
is an additional resonance at $\Delta = V/2$ in which we drive a four-photon
process from $\ket{g,g}$ to $\ket{r,r}$ through the off-resonant intermediate
state $\ket{W}$. Using spectroscopy, we determine this 4-photon resonance to be at
$\Delta \sim 2\pi \times 12.2$~MHz , from which we calculate $V = 2\Delta =
2\pi \times 24.4$~MHz. This is consistent with independent measurements of our
trap spacing of approximately $5.7~\mu$m, from which we additionally calibrate the
spacing used in other arrangements ($3.57~\mu$m for Z$_3$ order and
$2.87\,\mu$m for Z$_4$ order).

\noindent
\textbf{Timing limits imposed by turning off traps.} Atoms can be
unintentionally lost owing to motion away from the trapping region during the
Rydberg pulse when the traps are off. This process depends on the atomic
temperature and for how long we turn off the traps. In particular, with our
measured temperature of $12~\mu$K~(\ref{fig:ED2}), the loss due to atomic
motion for trap-off times of $< 4~\mu$s is only about $0.1\%$. For longer
trap-off times, we see loss of up to $2\%$ at $6~\mu$s or $9\%$ at $10~\mu$s.
To cap this infidelity at $3\%$, all experiments described in the main text and
Methods operate with trap-off times of $\le 7~\mu$s.

\noindent
\textbf{State detection fidelity.}
Each atom is identified as being in $\ket{g}$ (or $\ket{r}$) at the end of the
Rydberg pulse by whether it is (or is not) present in the third fluorescence
image. Detection infidelity arises from accidental loss of atoms in $\ket{g}$
or accidental recapture of atoms in $\ket{r}$. For an atom in state $\ket{g}$,
detection fidelity is set by the finite trap lifetime (which causes baseline
loss of $1\%$) and motion due to turning the traps off ($\le 3\%$ for all
experiments shown, see Methods section on `Timing limits'). For
the 7-atom data shown in Fig.~\ref{fig:preparation_fidelity} and the 51-atom data shown
in Figs~\ref{fig:size-scaling} and~\ref{fig:large_system_size}, we measured ground state detection fidelities of $98\%$ and
$99\%$, respectively.

For an atom in state $\ket{r}$, the optical tweezer yields an anti-trapping
potential, but there is a finite probability that the atom will decay back to
the ground state and be recaptured by the tweezer before it can escape the
trapping region. We quantify this probability by measuring Rabi oscillations
between $\ket{g}$ and $\ket{r}$~(\ref{fig:ED3}) and extracting the maximum
amplitude of the oscillation signal. After accounting for the loss of ground
state atoms as an offset to the signal, we obtain a typical effective detection
fidelity of $93\%$ for the $\ket{70S_{1/2}}$ Rydberg state. Furthermore, we
observe a reduced detection fidelity at lower-lying Rydberg states, which is
consistent with the dependence of the Rydberg lifetime on the principal quantum
number~\cite{Beterov2009}.

\noindent
\textbf{Correcting for finite detection fidelity.}
The number of domain walls is a metric for the quality of preparing the desired
crystal state. Boundary conditions make it favorable to excite the atoms at the
edges. Therefore, we define a domain wall as any instance where two
neighbouring atoms are found in the same state or an atom at the edge of the
array is found in state $\ket{g}$. In systems composed of an odd number of
particles, this definition sets the parity of domain walls to be even.

The appearance of domain walls can arise from non-adiabaticity across the phase
transition, as well as scattering from the intermediate $6P$ state, imperfect
optical pumping, atom loss or other processes (see Methods section ` Coherence
limitations'). However, the observed number of domain walls is increased artificially
owing to detection infidelity; any atom within a crystal domain that
is misidentified increases the number of measured domain walls by two. For this
reason, we use a maximum-likelihood routine to estimate the parent
distribution, which is the distribution of domain walls in the prepared state
that best predicts the measured distribution. We use two methods to correct for
detection infidelity, depending on whether we are interested in only the
probability of generating the many-body ground state or in the full probability
distribution of the number of domain walls.

\noindent
\textbf{Correcting detection infidelity.} \emph{Many-body ground-state
preparation.} Having prepared the many-body ground state, the probability of
correctly observing it depends on the measurement fidelity for atoms in the
electronic ground state $f_g$, the measurement fidelity for atoms in the
Rydberg state $f_{r}$, and the size of the system $N$. Assuming a perfect
crystal state in the Z$_{2}$ phase, the total number of atoms in the Rydberg
state is $n_{r} = (N+1)/2$, while the number of atoms in the ground state is
$n_{g} = (N-1)/2$.  The probability of measuring the perfect state is then $p_m
= f_{r}^{n_{r}}\times f_{g}^{n_{g}}$. Therefore, if we observe the ground state
with probability $p_{\rm exp}$, the probability of actually preparing this
state is inferred to be $p_{\rm exp}/p_{m}$. The blue data points in Fig.~\ref{fig:size-scaling}a
are calculated this way.

\noindent
\emph{Maximum likelihood state reconstruction.}
To correct for detection fidelity in the entire distribution of domain
walls, we use a maximum-likelihood protocol. For this purpose, we assume that
the density of domain walls is low, such that the probability of preparing two
overlapping domain walls, meaning three consecutive atoms in the same state, is
negligibly small. Under this assumption, misidentifying an atom within a domain
wall shifts its location, but does not change the total number. However,
misidentification of an atom within a crystal domain increases the number of
domain walls by two. For any prepared state with a number of domain walls
$n_i$, we can calculate the probability to measure $n_f$ domain walls,
$p(n_f|n_i)$. We can construct a matrix $M$, which transforms an initial
probability distribution for the number of domain walls, ${\bf W}_i = (p(n_i = 0),
p(n_i = 2), ...)$, into the expected observed distribution ${\bf W}_f = M {\bf
W}_i$, where $M_{kl} = p(n_f=k|n_i=l)$.  Given an experimentally observed
distribution of domain walls, ${\bf W}_o$, and a test initial distribution
${\bf W}_i'$, we can calculate the difference vector between them ${\bf D}' =
{\bf W}_o - {\bf W}_f' = {\bf W}_o - M {\bf W}_i'$.

Using ${\bf D}'$ and the 68\% confidence intervals of the measured data
($\boldsymbol{\sigma}$), obtained via an approximate parametric bootstrap
method~\cite{Glaz1999}, we define a cost function

\begin{equation}
    \label{eq:CostFunction}
    C\Big({\bf W}_o, {\bf W}_i'\Big) = \sum_{k} \bigg(\frac{D'_{k}}{\sigma_{k}}\bigg)^{2},
\end{equation}
where the sum is taken over the elements of the vectors. We find the most
likely parent distribution ${\bf W}_i$ by minimizing the cost function over
the different possible ${\bf W}_i'$, under the constraints that that every
element is between 0 and 1, and the sum of the elements is 1. For this purpose,
we use a sequential least-square programming routine. To reduce biases, we use
a random vector as a starting point of the minimization procedure. We checked
that repeating the procedure several times with different initial vectors
converged to the same parent distribution, and that the distribution of domain
walls predicted by this parent distribution was in excellent agreement with the
measured distribution. The result of such a procedure on the dataset used for
Fig.~\ref{fig:large_system_size}c is shown in~\ref{fig:ED7}.

\noindent
\textbf{Adiabatic pulse optimization.}
To prepare the ordered phases, we use frequency chirped pulses by
varying the two-photon detuning $\Delta$ across the bare $\ket{g}
\leftrightarrow \ket{r}$ resonance, corresponding to $\Delta=0$. To perform
these sweeps, we drive a high-modulation-bandwidth voltage-controlled
oscillator (Mini-Circuits ZX95-850W-S+) according to either cubic or
tangent functional forms:

\begin{equation}
\begin{aligned}
    \label{eq:CubicTangentSweep}
    &V(t)_{\rm cubic} = a(t-t_0)^3 + b(t-t_0) + c\Big|_{\Delta_{\rm min}\leq\Delta\leq\Delta_{\rm max}}
\\
&V(t)_{\rm tangent} = a \tan\left( b(t-t_0) \right) + c\Big|_{\Delta_{\rm min}\leq\Delta\leq\Delta_{\rm max}}
\end{aligned}
\end{equation}
with programmable parameters $a, b, c$. The output from this voltage-controlled
oscillator is mixed (Mini-Circuits ZFM-2-S+) with a $750\,$MHz source to
generate the difference frequency, which is used to drive the acousto-optic modulator in the
$420$-nm-light path. The detuning $\Delta$ is set to truncate at minimum and
maximum values $\Delta_{\rm min}$ and $\Delta_{\rm max}$, respectively. The
tangent adiabatic sweep was used for datasets with 51~atoms (Figs~\ref{fig:size-scaling} and~\ref{fig:large_system_size})
owing to improved performance, whereas the cubic
form was used for all smaller system sizes and for the data on crystal
dynamics (Fig.~\ref{fig:crystal_dynamics}).

At the end of the sweep, the number of domain walls in the crystal provides a
metric for the quality of the crystal preparation. All parameters in
equation~\eqref{eq:CubicTangentSweep} are iteratively optimized as to minimize
the domain wall number, or equivalently, to maximize the crystal preparation
fidelity. The optimization starts with the offset $c$, followed by the
parameter $b$, the maximum and minimum detunings $\Delta_{\rm min/max}$,
and finally the parameter $a$. Repeated optimization of these parameters often
leads to better crystal preparation fidelities~\cite{Johansson2012}.

After passing through the acousto-optic modulator, the $420$-nm light is
coupled into a fibre. The coupling is optimized for the voltage-controlled
oscillator frequency at which the light is resonant with the $\ket{g} \to
\ket{r}$ transition ($f_{\rm opt}$), and decreases as the voltage-controlled
oscillator frequency deviates from $f_{\rm opt}$. The power throughout all
frequency sweeps is $\geq 75$\% of the power at $f_{\rm opt}$.

\noindent
\textbf{Coherence limitations.}
When sweeping into the crystalline phase, the control parameter $\Delta(t)$
must be varied slowly enough that the adiabaticity criterion is sufficiently
met. However, for long pulses, additional technical errors may become limiting.
Here, we summarize some key limitations:
\begin{itemize}
  \item \textbf{State preparation fidelity:} For all data analysed, we
      preselect defect-free atom arrays. The preparation fidelity is therefore
      given by the probability that each atom in the array is still present for
      the Rydberg pulse, and that it is prepared in the correct magnetic
      sublevel: $\ket{5S_{1/2}, F=2, m_F = -2}$. Including both factors, we
      estimate that atoms are present and in the correct magnetic sublevel with
      fidelity $f > 98\%$. For experiments with 51~atoms, this leads to
      at most about one atom prepared incorrectly .

\item \textbf{Spontaneous emission:} The $70S$ Rydberg state has an estimated
    lifetime of $150\,\mu$s (including blackbody radiation at
    $300\,$K)~\cite{Beterov2009}. In additional, for the typical intermediate
    detuning $\Delta \approx 2\pi \times 560\,$MHz and the single photon infrared and
    blue Rabi frequencies of $(\Omega_R, \Omega_B) \approx 2\pi \times (36,
    60)\,$MHz, spontaneous emission from the intermediate state occurs on a
    timescale of $40\,\mu$s for the ground state, and introduces a combined
    effective lifetime of $50\,\mu$s for the Rydberg state. This leads to an
    average scattering rate of $2\pi\times3.6\,$kHz.

  \item \textbf{Rabi frequency homogeneity:} We aim to align our beams to
      globally address all trapped atoms with a uniform Rabi frequency $\lvert
      \Omega_i \rvert= \Omega$. Experimentally, we achieve homogeneity up to
      differences of about $3\%$~(\ref{fig:ED3}b).

  \item \textbf{Intensity fluctuations:} Primarily because of pointing
      instability, the global Rabi frequency fluctuates by small amounts from
      shot to shot of the experiment. To reduce slow drifts of the
      beams, we use a 1:1.25 telescope to image on a camera their position on
      the plane of the atoms and feedback to stabilize their position to a
      target every 500 repetitions (about 2 minutes).

  \item \textbf{Rydberg laser noise:} The coherence properties of the Rydberg
      lasers over typical experimental times are probed by measurements on
      single, non-interacting atoms. In particular, spin echo measurements
      between $\ket{g}$ and $\ket{r}$ show no visible decay of coherence over
      $5\,\mu$s~(\ref{fig:ED3}c). This measurement, along with the measured
      noise contribution from the laser lock of $< 0.5$~kHz (see Methods section
      `Rydberg lasers set-up'), indicates that the line widths of the laser are
      sufficiently narrow. Additional phase noise is introduced by the laser
      lock around the lock bandwidth of about $1\,$MHz. This phase noise may
      cause weak additional decoherence on the adiabatic sweep experiments
      shown in the main text.

  \item \textbf{Finite atomic temperature:} Our finite atomic temperature of
      approximately $12\,\mu$K introduces both random Doppler shifts (of about
      $2\pi \times 50\,$kHz) and fluctuations in the atomic positions (about
      $120\,$nm radially, $600\,$nm longitudinally) for each atom in each cycle
      of the experiment. The Doppler shift is very small in magnitude compared
      to the single atom Rabi frequency $\Omega$. The position fluctuations can
      introduce noticeable fluctuations in the interaction energy between a
      pair of atoms from shot to shot. As an example, at our chosen lattice
      spacing of $5.9\,\mu$m, we calculate an interaction energy of $2\pi
      \times 24\,$MHz. However, if the distance fluctuates by about
      $\sqrt{2}\times120\,$nm $\approx 170\,$nm, then the actual interaction
      energy can range from $2\pi\times 21\,$MHz to $2\pi\times29\,$MHz. The
      longitudinal position fluctuations add in quadrature, so they contribute
      less to fluctuations in distance.

  \item \textbf{Electric and magnetic fields:} We observed that the
      Rydberg resonance can drift over time, especially for states with high
      principal quantum number $n$, which we attribute to uncontrolled
      fluctuations in the electric field. We can reduce these fluctuations by
      shining $365$-nm ultraviolet light on the glass cell in between experimental
      sequences and during the magneto-optical trap loading period, which
      stabilizes the charge environment on the glass cell surface. While the
      fluctuations for states $n \geq 100$ are still significant, they become
      negligible ($< 100\,$kHz) for our chosen state $n=70$.
    
The energy shifts of the initial state $\ket{g}$ and final state $\ket{r}$ with
magnetic fields are identical. Differential shifts of the intermediate state
are very small compared to the detunings of the two laser beams from the
$6P_{3/2}$ state. Therefore, we do not expect magnetic fields to play any
significant role in fluctuations between experimental runs.
\end{itemize}

We note that the use of deterministically prepared arrays allows us to
optimize the coherence properties efficiently. For example, for collective Rabi
oscillations of fully blockaded groups of up to three atoms, we observe an
improvement in the product $\Omega\tau_d$ of about an order of magnitude
compared to previous work~\cite{Labuhn2016}, where $\tau_d$ is the decay time
of the Rabi oscillations. In addition, the relatively high fidelity in the
preparation of Z$_2$ ordered states with 51 atoms (\ref{fig:ED5}) indicates a
significant amount of coherence preserved over the entire evolution. These
considerations indicate that the present approach is promising for near-term
coherent experiments with large scale systems~\cite{Boixo2016}.

\noindent
\textbf{Comparison with a classical thermal state.}
To gain some insight into the states obtained from our preparation protocol
(Fig.~\ref{fig:preparation_fidelity}a), we provide a quantitative comparison between
experimentally measured quantities and those computed from a thermal ensemble.
In particular, we note that, deep in the ordered phase $\Delta/\Omega \gg 1$,
the coherent coupling of the ground state to the Rydberg state can be neglected
owing to strong energetic suppression and that the effective Hamiltonian
becomes diagonal in the measurement basis.  This allows us to calculate all
properties of a thermal state even for systems of 51 atoms by
computing the partition function explicitly via the transfer matrix
method~\cite{Baxter2007}.  Also, we may consider the interactions only up to
next-nearest neighbours because the coupling strengths for longer distances are weak
compared to the maximum timescale that is accessible in our experiments.  To this end,
we consider the Hamiltonian 
\begin{equation*}
    \ham_{\rm cl}=-\Delta\sum_{i=1}^N
n_i+\sum_{i=1}^{N-1}V_1 n_i n_{i+1}+\sum_{i=1}^{N-2}V_2 n_i n_{i+2}
\end{equation*}
The eigenstates of this Hamiltonian are simply $2^N$ classical configurations,
where each atom is in either $\ket{g}$ or $\ket{r}$. We label these
configurations by a length-$N$ vector ${\bf i}=(i_1,i_2,\dots,i_N)$
($i_n\in\{g,r\}$), and denote their energy by $E_{\bf i}$. In a thermal
ensemble $\rho=\exp(-\beta \ham_{\rm cl})/Z$ with $Z \equiv
\textrm{tr}[\exp(-\beta \ham_{\rm cl})]$ and inverse temperature $\beta$, the
probability to find a particular configuration ${\bf i}$ is $p_{\bf
i}=\exp(-\beta E_{\bf i})/Z$. Because $E_{\bf i}$ can be written as a sum of
local terms involving interactions only up to a range of two, the partition sum can
be evaluated using a standard transfer matrix of size $4\times 4$.  Moreover,
using this approach, we can evaluate all measurable quantities for the thermal
ensemble, including the average number of domain walls $\mean{D}=\tr{D\rho}$,
where 
\begin{equation*}
D=\sum_{i=1}^{N-1}( n_i n_{i+1}+(1-n_i)(1- n_{i+1}))+(1-n_1)+(1-n_N)
\end{equation*}
is an operator counting the number of domain walls, the
correlation function
\begin{equation*}
    g^{(2)}(d) = \frac{1}{N-d} \sum_{i=1}^{N-d} g^{(2)}_{i,i+d}
\end{equation*}
and even the full counting statistics for the domain wall distribution in the
state $\rho$.  In particular, the probability to measure exactly $n$ domain
walls $p_n=\tr{P_n\rho}$ can be computed from a Fourier transform of the
Kronecker delta function
\begin{equation*}
    P_n\equiv \delta_{D,n} =
\frac{1}{N+2}\sum_{k=0}^{N+1}\exp\left[i\frac{2\pi}{N+2}k(n-D)\right]
\end{equation*}
with $n=0,1,2,\dots N+1$.

We can include the effect of imperfect detections in this formalism. To that
end, we denote the expectation value of an observable $O$ as
\begin{align}\label{eq:thermal_infidel}
    \meanmean{O}=\sum_{{\bf i},{\bf j}} O_{\bf i} \Lambda_{{\bf i},{\bf j}} p_{\bf j},
\end{align}
where $O_{\bf i}$ is the value of the observable in state ${\bf i}$, and
$\Lambda_{{\bf i},{\bf j}}$ is the probability of detecting state ${\bf i}$ when
the system is in state ${\bf j}$, accounting for finite detection fidelity.
Assuming detection errors occur independently from one another, we have
\begin{equation*}
\Lambda_{{\bf i},{\bf j}}=\prod_n \lambda_{i_n,j_n}
\end{equation*}
where $\lambda_{g,g}=f_g$ is the probability to correctly detect an atom in the
ground state, $\lambda_{r,r}=f_r$ is the probability to correctly detect an
atom in the Rydberg state, and $\lambda_{r,g}=1-\lambda_{g,g}$, and
$\lambda_{g,r}=1-\lambda_{r,r}$. Equation~\eqref{eq:thermal_infidel} can be
evaluated using a $16\times 16$ transfer matrix for any observables of
interest.

To obtain a quantitative comparison with our experiments, we determine the
inverse temperature $\beta$ in such a way that the average number of domain
walls, including the effect of imperfect detections, matches the
experimentally determined value, $\meanmean{D}=9.01(2)$. For
$\Delta=2\pi\times 14\,$MHz, $V_1=2\pi\times 24\,$MHz and $V_2=2\pi\times
0.38\,$MHz, this leads to $\beta=3.44(1)/\Delta$ or equivalently to the entropy
per atom of $s/k_B=0.286(1)$~(\ref{fig:ED8}a,b). Because $\beta$ characterizes
the thermal state completely, we can extract the corresponding domain-wall
distribution~(\ref{fig:ED8}c) and the correlation function~(\ref{fig:ED8}d) as
described above. We find that the correlation length in the corresponding
thermal state is  $\xi_{\rm th}= 4.48(3)$, which is significantly longer than
the measured correlation length $\xi=3.03(6)$, from which we deduce that the
experimentally prepared state is not thermal.

\noindent
\textbf{Dynamics after sudden quench.}
To understand the dynamics of the Z$_2$ Rydberg crystal after quenching
the detuning to $\Delta=0$, we first consider a simplified model, in which
interactions beyond nearest neighbour are neglected. In addition, we replace
the nearest neighbour interactions with the hard constraint that two neighbouring
atoms cannot be excited at the same time. Such an approximation is well
controlled in the limit of $V_{i,i+1} \gg \Omega$, as in the case of our
experiments, for a time exponentially long in
$V_{i,i+1}/\Omega$~\cite{Abanin2016}. In this limit, the Hamiltonian is
approximated by 
\begin{equation*}
\ham_{\rm c}=\sum_{i}P_g^{i-1}\left(\frac{\Omega}{2}\sigma_x^i-\Delta P_r^i\right)P_g^{i+1},
\end{equation*}
where $P_g^i=\ketbra{g_i}{g_i}$, $P_r^i=\ketbra{r_i}{r_i}$. We identify
$P_g^{i=0}=P_g^{i=N+1}=1$ at the boundaries. Within this approximation, the
relevant Hilbert space consists only of states with no neighbouring atoms in
the Rydberg state, i.e. $P_r^{i}P_r^{i+1}=0$. The dimension of this constrained
Hilbert space is still exponentially large and grows as $\sim \phi^N$, where
$\phi= 1.618\dots$ is the golden ratio. 

In the simplest approximation, we can treat the array of atoms as a collection
of independent dimers,  $\ket{\Psi(t)}=\bigotimes_i\ket{\phi(t)}_{2i-1,2i}$,
where for each pair of atoms only three states are allowed owing to the blockade
constraint: $\ket{r,g}$, $\ket{g,g}$ and $\ket{g,r}$. The dynamics of each pair
with initial state $\ket{\phi(0)}=\ket{r,g}$ is then
\begin{equation*}
    \begin{aligned}
        \ket{\phi(t)}=&\frac{1}{2}\left[1+\cos(\Omega
t/\sqrt{2})\right]\ket{r,g}+\frac{i}{\sqrt{2}}\sin(\Omega
t/\sqrt{2})\ket{g,g}\\
&+\frac{1}{2}\left[1-\cos(\Omega t/\sqrt{2})\right]\ket{g,r}
\end{aligned}
\end{equation*}
This dimer model predicts that each atom flips its state with respect to its initial
configuration after a time $\tau=\sqrt{2}\pi/\Omega$. The corresponding
oscillations between two complementary crystal configurations are thus a factor
$\sqrt{2}$ slower than an independent spin model would predict, which is
qualitatively consistent with the experimental observations. We note that this
dimerized ansatz does not satisfy the constraint $P_r^{i}P_r^{i+1}=0$ between
two neighbouring dimers, which is an artefact originating from the artificial
partitioning of the array into non-interacting dimers.

To go beyond this approximation, we consider an ansatz for the many-body
wavefunction that treats each atom on an equal footing. The simplest such
wavefunction that also allows for non-trivial entanglement between the atoms
can be written as a matrix product state with bond dimension
2~\cite{Schollwock2011}.  In particular we consider a manifold of states of the
form 
\begin{equation*}
    \ket{\Psi(\{\theta_n\})}=\sum_{\{i_n\}}v_L
A(\theta_1)^{i_1}A(\theta_2)^{i_2}\cdots A(\theta_N)^{i_N}
v_R\ket{i_1,i_2,\dots,i_N}
\end{equation*}
with matrices 
\begin{equation*}
A(\theta_n)^{g}=\left(\begin{array}{cc}\cos(\theta_n) & 0 \\ 1 &
    0\end{array}\right),\quad  A(\theta_n)^{r}=\left(\begin{array}{cc}0 &
    i\sin(\theta_n) \\ 0 & 0\end{array}\right)
\end{equation*}
and boundary vectors $\quad v_{L}=\left(\begin{array}{cc}1, &
1\end{array}\right)$ and  $v_{R}=\left(\begin{array}{cc}1, &
0\end{array}\right)^\intercal$.  Here, the indices $i_n\in \{g,r\}$
enumerate the state of the $n$th atom. This manifold satisfies the
constraint that no two neighbouring atoms are excited simultaneously. The
many-body state within this subspace is completely specified by the $N$
parameters $\theta_n\in [0,2\pi]$.  In particular, it enables the initial
crystal state to be represented
by $\theta_{2n-1}=\pi/2$ for atoms on odd sites and
$\theta_{2n}=0$ for atoms on even sites, as well as its inverted version,
$\theta_{2n-1}=0$ for odd and $\theta_{2n}=\pi/2$ for even sites,
respectively.  Using the time-dependent variational
principle~\cite{Haegeman2011}, we derive equations of motion for the wave
function within this manifold. For an infinite system with a staggered
initial state $\theta_{n+2}=\theta_{n}$, such as the Z$_2$-ordered state,
the wave function is at all times described by two parameters
$\theta_a=\theta_{2n-1}$ and $\theta_{b}=\theta_{2n}$ for even and odd
sites. The corresponding non-linear, coupled equations of motion are
\begin{equation}
    \label{MPS_EOM}
    \begin{aligned}
\dot\theta_a&=-\frac{1}{2} \sec \left(\theta _b\right) \left[\sin \left(\theta _a\right) \cos
   ^2\left(\theta _a\right) \sin \left(\theta _b\right)+\cos ^2\left(\theta
   _b\right)\right]\\
   \dot\theta_b&=-\frac{1}{2} \sec \left(\theta _a\right) \left[\sin \left(\theta _b\right) \cos
   ^2\left(\theta _b\right) \sin \left(\theta _a\right)+\cos ^2\left(\theta
   _a\right)\right]
   \end{aligned}
   \end{equation}
A numerical solution of these variational equations for the crystalline initial
state predicts a periodic motion with a frequency of approximately
$\Omega/1.51$~(\ref{fig:ED9}), with the many-body wavefunction
oscillatin between two staggered configurations.  

\noindent
\textbf{Decay of the oscillations and growth of entanglement after the quantum quench.}
To obtain more insight into the dynamics of our system beyond these
variational models, we use exact numerical simulations to integrate the
many-body Schr\"odinger equation. In particular, we focus on the decay of
oscillations and the growth of entanglement entropy in our system. Owing to the
exponentially growing Hilbert space, this method is limited to relatively small
system sizes.  We make use of the constrained size of the Hilbert space
(blockade of nearest-neighbouring excitations of Rydberg states), and propagate
the state vector of up to 25 spins using a Krylov subspace projection method.
In~\ref{fig:ED10}a we show the dynamics of the domain wall density under the
time evolution of the constrained Hamiltonian $\ham_c$ with $\Omega=2\pi\times
2 \,$MHz and $\Delta=0$.  We consider two different initial states: the
disordered state in which each atom is initially prepared in the ground state
$\ket{g}$, and the perfect crystalline state $\ket{r,g,r,g,\dots}$. We note
that in both cases the energy density corresponds to that of an
infinite-temperature thermal ensemble in the constrained subspace with respect
to $\ham_c$.

For the disordered initial state, the domain wall density relaxes quickly to a
steady state value. In contrast, if the system is initialized in the perfect
crystalline state, the domain wall density oscillates for long times and decays
at a rate much slower than the oscillation period. We confirmed numerically
that this initial decay time is independent of the system size. We further note
that for every system size accessible in our numerical method, the domain wall
density does not relax to a steady value even at very long times, but continues
to oscillate with a reduced amplitude. Moreover, whereas the disordered initial
state relaxes to an average domain wall density that is consistent with a thermal state
of infinite temperature corresponding to the energy density of the initial
state, this is clearly not the case for the crystalline initial state. This
qualitatively distinct behavior for two different initial states is also
reflected in the growth of entanglement entropy after the
quench~(\ref{fig:ED10}c, dashed lines). Although in both cases the entanglement
entropy grows initially linearly, the rate of growth is significantly lower for
the crystalline initial state. Moreover, unlike the case of disordered initial
state, in which the entanglement entropy saturates quickly to its maximum value
(limited by the finite system size and the constrained Hilbert space), for the
crystalline initial state the entanglement entropy does not seem to approach
the same value.

To understand the influence of the $1/R^6$-decaying interactions, we show the
corresponding dynamics and entanglement growth in~\ref{fig:ED10}b, c (solid
lines). Numerically, we treat the strong nearest neighbour interactions
perturbatively -- by adiabatic eliminations of simultaneous excitation of
neighbouring Rydberg states -- and the weak interactions beyond nearest
neighbours exactly. For the disordered initial state, we find that
the dynamics of domain wall density and the entanglement growth remain similar
to the previous case, in which long-range interactions are neglected; in this
case, the thermalization time is barely affected.  In contrast, for the
crystalline initial state, the oscillations decay significantly faster when
next-to-nearest neighbour interactions are included. We therefore attribute the
thermalization in this case to interactions beyond the nearest-neighbour
blockade constraint. From the growth of the entanglement entropy we see that
the crystalline initial state still thermalizes slower than the disordered
initial state.

\noindent
\textbf{Numerical time evolution via matrix product state algorithm.}
The numerical data presented in Figs~\ref{fig:large_system_size}b
and~\ref{fig:crystal_dynamics}b were obtained by simulating the evolution of
the 51-atom array during the sweep across the phase transition and the
subsequent sudden quench using a matrix product state algorithm with bond
dimension $D=256$. We simulate the entire preparation protocol to generate the
Rydberg crystal (Fig.~\ref{fig:large_system_size}b), and use the resulting
state as an initial state for the time evolution after the sudden quench. To
this end, we use a time-evolving block decimation
algorithm~\cite{Vidal2004,Daley2004}, with a Suzuki-Trotter splitting of the
Hamiltonian to update the state. The time step used in this Trotterization is
$\Omega \Delta t=0.004$. We take into account only nearest-neighbour and
next-nearest-neighbour interactions, neglecting small interactions for atoms
that are separated by three or more sites (as discussed also in Methods section
`Comparison with a classical thermal state'). We account for finite detection
fidelities that are determined independently, but otherwise do not include any
incoherent mechanisms.  Remarkably, for local quantities, such as the
domain-wall density, this fully coherent simulation agrees well with the
experimentally measured values. For higher-order correlation functions, such as
the variance of the number of domain walls, the fully coherent simulation and
the experiment agree only qualitatively~(\ref{fig:ED6}). The quantitative
difference is probably due to either limitations of the MPS simulations or
various incoherent processes being present in the experiment.

\FloatBarrier
\textbf{Data availability.} The data that support the findings of this study
are available from the corresponding author upon reasonable request.
\renewcommand{\addcontentsline}[3]{}

\bibliographystyle{h-physrev}

\begin{figure*}
    \includegraphics[width = 1.0 \textwidth]{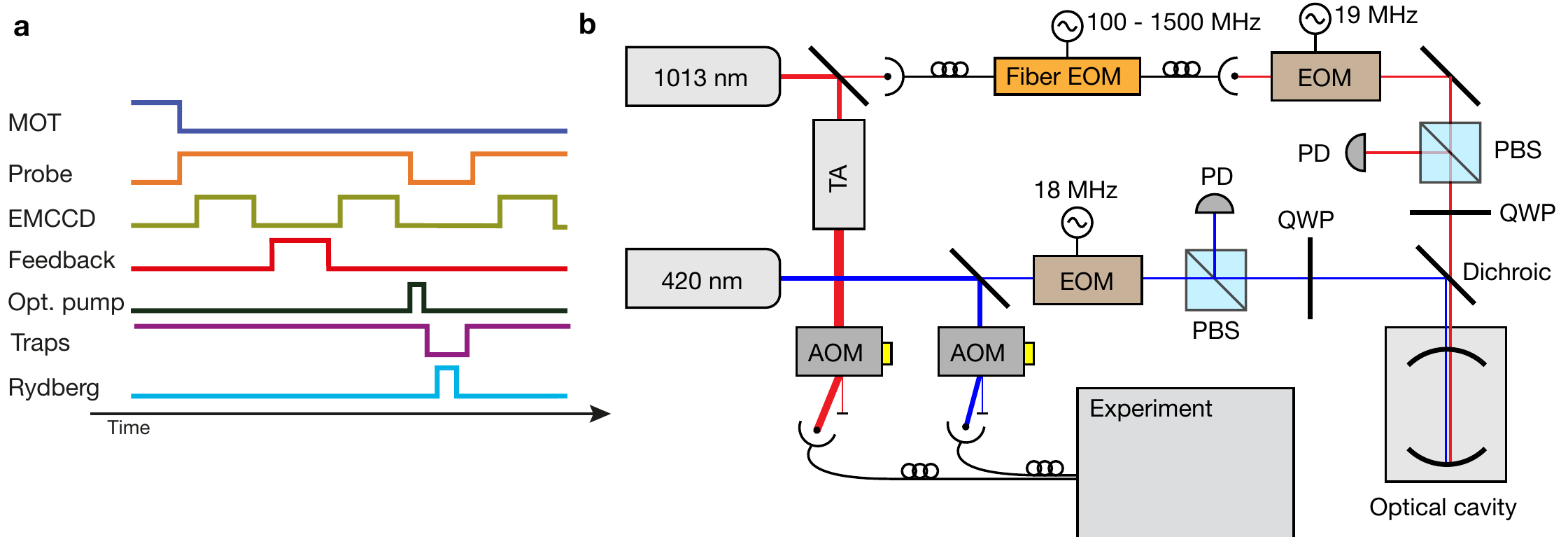}
    \caption{\textbf{Experimental sequence and Rydberg laser set-up.}
    \textbf{a,} The tweezer array is initially loaded from a magneto-optical
    trap. A single-site-resolved fluorescence image taken with an
    electron-multiplying CCD camera (EMCCD) is used to identify the loaded
    traps. Using this information, a feedback protocol rearranges the loaded
    atoms into a preprogrammed configuration, which is verified by the second
    EMCCD image. After that, all atoms are optically pumped into the $\ket{F=2,
    m_F = -2}$ state, the tweezers are turned off, and the Rydberg lasers are
    pulsed. After the traps are turned back on, a third EMCCD image is taken to
    detect Rydberg excitations with single-site resolution. \textbf{b,}
    Schematic representation of the Rydberg laser set-up, which is used to
    stabilize two external cavity diode lasers to a reference optical cavity
    with a fast Pound-Drever-Hall lock. Key: TA, tapered amplifier; AOM,
    acousto-optic modulator; EOM, electo-optic modulator; PD, photodetector;
    PBS, Polarizing beam splitter; QWP, quarter-wave plate.\\ \\
\\ \\ \\ \\ \\ \\ \\ \\ \\ \\ \\ \\ \\ \\ \\}
    \label{fig:ED1}
\end{figure*}

\begin{figure*}
    \includegraphics{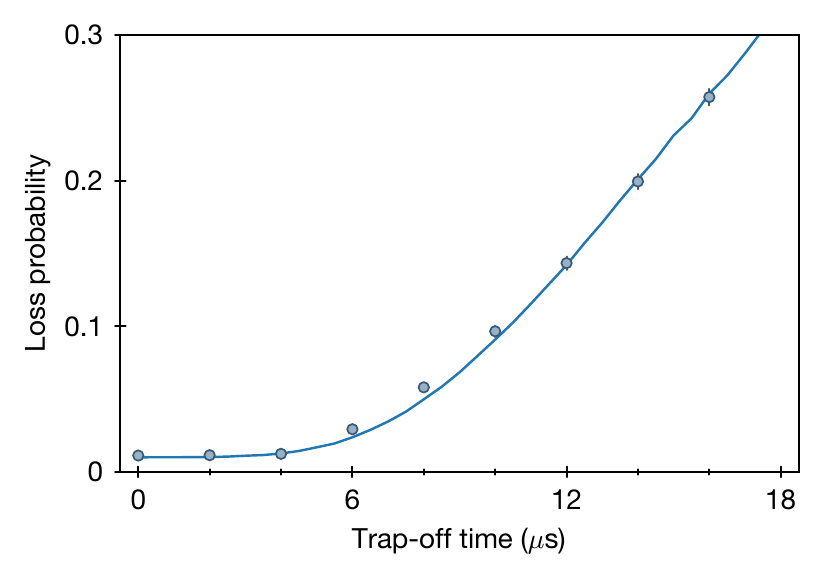}
  \caption{\textbf{Drop-recapture curve.} Measurements of atom loss probability
  as a function of trap-off time. For short times of up to $4\,\mu$s, the loss
  is dominated by finite trap lifetime ($1\%$ plateau). At larger trap-off
  times, the atomic motion away from the tweezer introduces additional losses.
  The solid line is a Monte Carlo simulation for a temperature of $11.8\,\mu$K.
  \\ \\ \\ \\ \\ \\ \\ \\ \\ \\ \\ \\ \\ \\ \\ \\ \\ \\ \\ \\ \\ \\ \\ \\ \\ \\
  \\ \\ \\ \\ \\}
  \label{fig:ED2}
\end{figure*}

\begin{figure*}
    \includegraphics[width = 1.0 \textwidth]{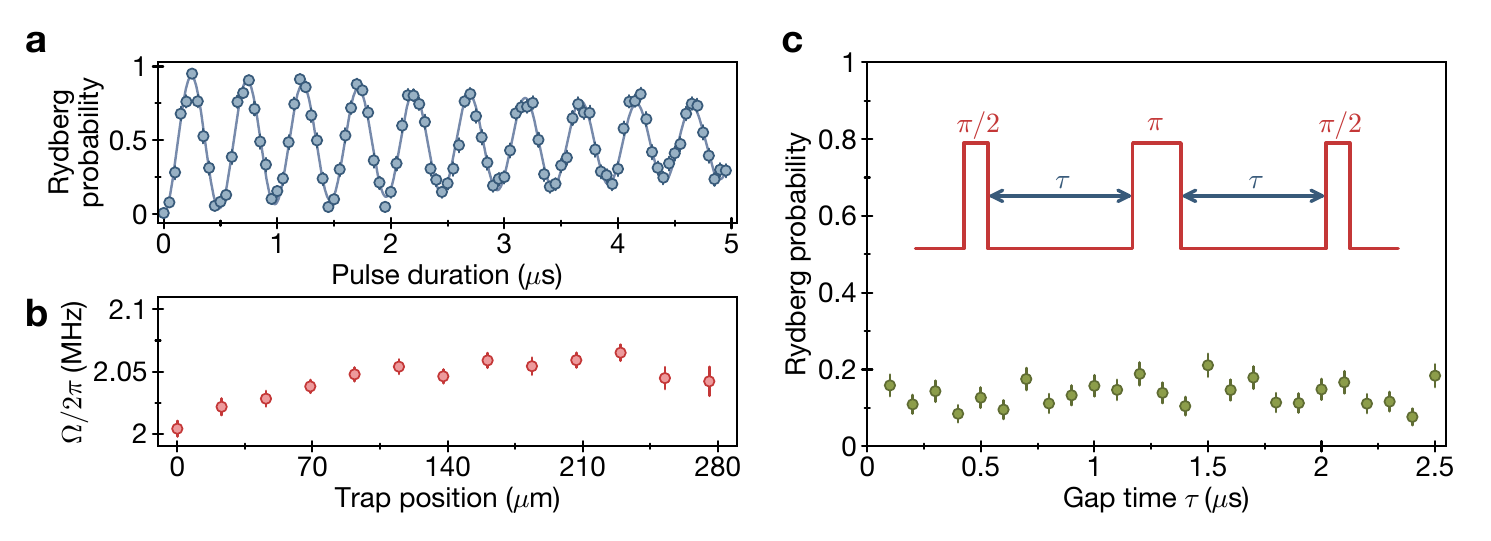}
    \caption{\textbf{Typical Rabi oscillation, homogeneity and coherence for
    non-interacting atoms} $a = 23\,\mu$m, $\Omega \gg V_{i,
    i+1} \approx5\,$kHz). \textbf{a,} Rabi oscillations. We observe a typical
decay time of about $6\,\mu$s, which is limited mainly by intensity fluctuations
from shot to shot. \textbf{b,} The fitted Rabi frequency for each atom across
the array (spatial extent of about $300\,\mu$m) is homogeneous to within $< 3\%$.
\textbf{c,} Measurement of the population in the Rydberg state after a spin
echo pulse sequence (inset). We find no decay of coherence over typical
measurement periods of several microseconds, thereby ruling out fast sources of
decoherence. Error bars in \textbf{a}-\textbf{c} denote 68\% confidence intervals.\\ \\ \\ \\ \\ \\
\\ \\ \\ \\ \\ \\ \\ \\ \\ \\ \\}
    \label{fig:ED3}
\end{figure*}

\begin{figure*}
    \includegraphics{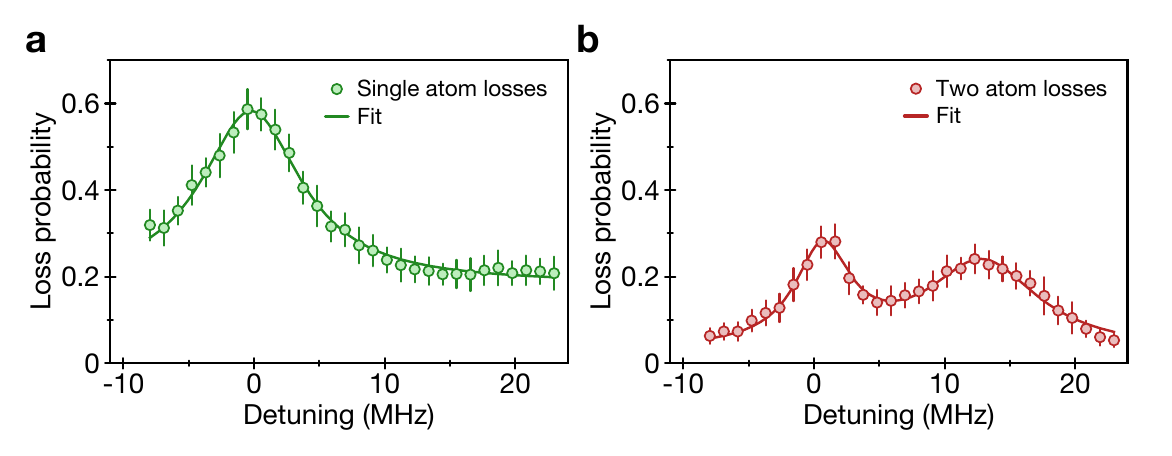}
  \caption{\textbf{Spectroscopic measurement of Rydberg interactions.}
  Spectroscopy on pairs of atoms separated by approximately $5.74~\mu$m.
  \textbf{a,} For single-atom losses, we observe a single peak at $\Delta = 0$
  corresponding to the two-photon coupling from $\ket{g,g}$ to $\ket{W}$.
  \textbf{b,} For two-atom losses, we observe an additional peak at $\Delta =
  2\pi \times 12.2$~MHz. This corresponds to the four-photon coupling from
  $\ket{g,g}$ to $\ket{r,r}$ through the intermediate state $\ket{W}$, detuned
  by $\Delta$. The interaction energy is then $V = 2\Delta$.
  This four-photon resonance is broadened as a result of random atom positions
  within the optical tweezers that result in fluctuations in interaction
  strengths from shot to shot of the experiment. Solid lines are fits to a
  single Lorentzian (\textbf{a}) and the sum of two Lorentzians (\textbf{b}).
  \\ \\ \\ \\ \\ \\ \\ \\ \\ \\ \\ \\ \\ \\ \\ \\ \\ \\ \\ \\ \\
  \\ \\ \\ \\ \\ \\ \\
  \\ \\ \\}
  \label{fig:ED4}
\end{figure*}

\begin{figure*}
    \includegraphics[width = 0.45 \textwidth]{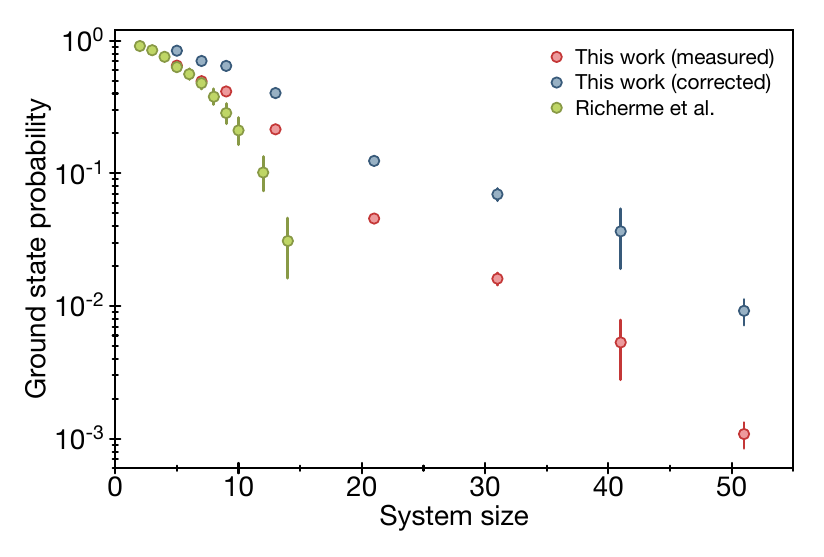}
  \caption{\textbf{Ground state preparation probability comparison.}
  We compared the ground state preparation probability obtained here
  (measured, red circles; corrected for detection infidelity, blue circles)
  with the most complete prior observations of a Z$_2$-symmetry breaking
  transition in a system of trapped ions (green circles)~\cite{Richerme2013}.
  We emphasize that the interaction Hamiltonians for the two systems are not
  identical, owing to the finite interaction range. In particular, the long range
  interactions tend to frustrate adiabatic transitions into Z$_2$ ordered
  states in~\cite{Richerme2013} and, to lesser extent, in this work. Error bars
  denote 68\% confidence intervals.}
  \label{fig:ED5}
\end{figure*}

\begin{figure*}
    \includegraphics[width = 0.8 \textwidth]{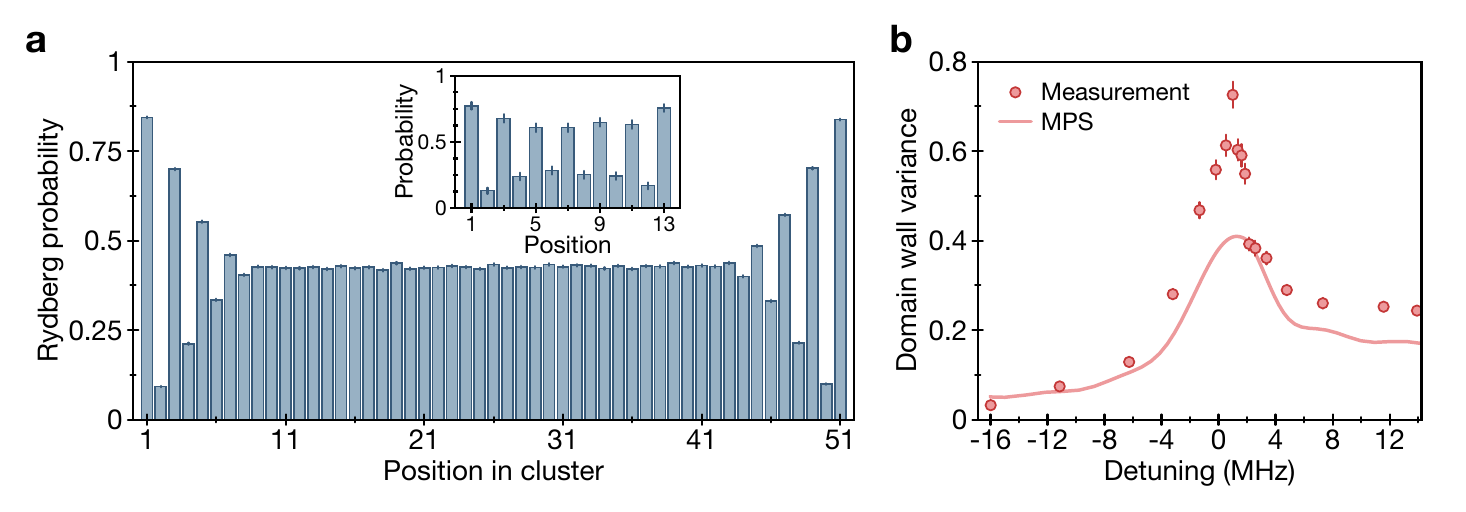}
  \caption{\textbf{State preparation with 51 atom clusters.} \textbf{a,}
  Average position-dependent Rydberg probability in a 51-atom cluster after the
  adiabatic sweep. The Z$_2$ order is visible at the edges of the system, while
  the presence of domain walls leads to an apparently featureless bulk
  throughout the center of the system. Inset, average Rydberg
  probabilities in a 13-atom chain, in which the Z$_2$ order is visible throughout
  the system, but the small system size prevents the study of bulk properties.
  \textbf{b,} Variance of the domain wall distribution during Z$_2$ state
  preparation. Points and error bars represent measured values. The solid red
  line corresponds to a full numerical simulation of the dynamics using a
  matrix product state ansatz (see text and Fig.~5). Error bars in \textbf{a} and
  \textbf{b} denote 68\% confidence intervals.\\ \\ \\
  \\ \\ \\ \\ \\ \\ \\ \\ \\ \\ \\ \\ \\ \\ \\ \\ \\ \\ \\ \\ \\ \\ \\ \\ \\ \\
  \\ \\ \\ \\ \\}
  \label{fig:ED6}
\end{figure*}

\begin{figure*}
    \includegraphics[width = 0.7 \textwidth]{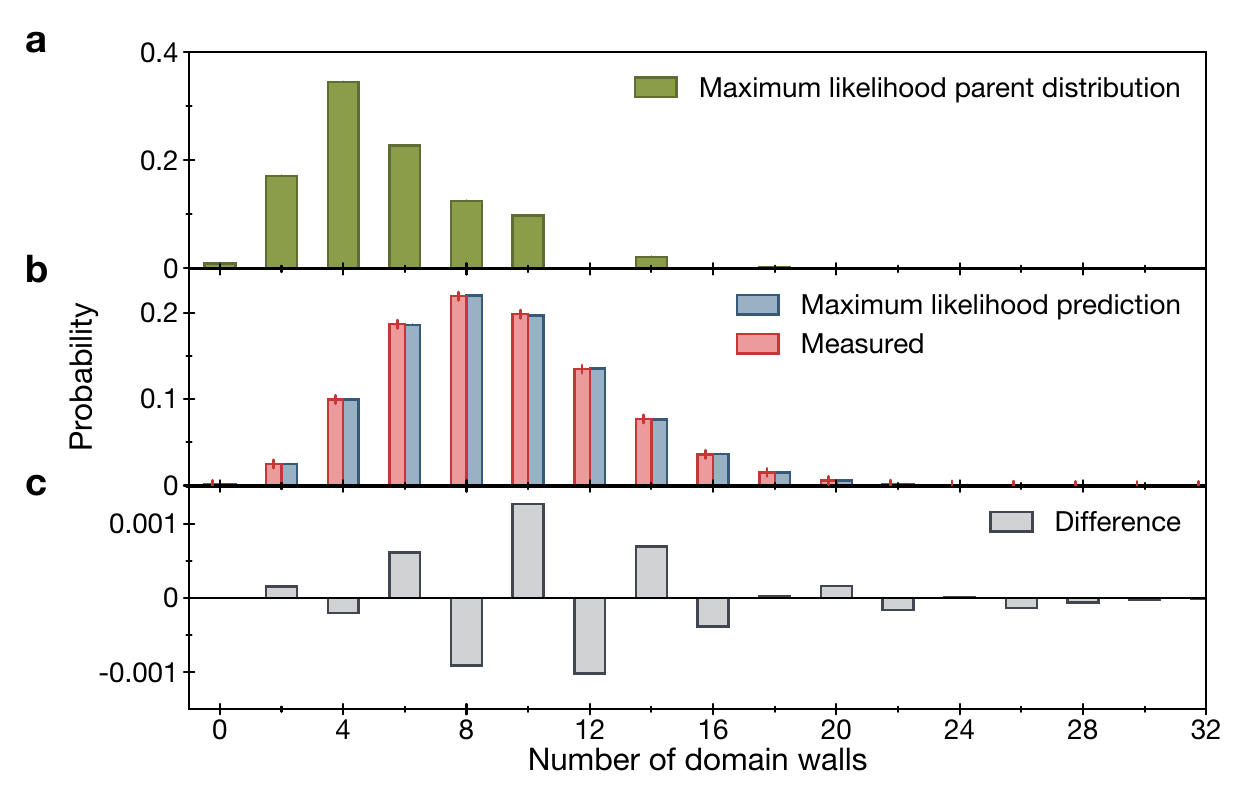}
    \caption{\textbf{State reconstruction.} \textbf{a,} Reconstructed parent
    distribution. \textbf{b,} Comparison of measured domain-wall distribution
    (red) and predicted observation given the parent distribution in \textbf{a}
    (blue). \textbf{c,} Difference between the two distributions in \textbf{b}. \\
    \\ \\ \\ \\ \\ \\ \\ \\ \\ \\ \\ \\ \\ \\ \\ \\}
    \label{fig:ED7}
\end{figure*}

\begin{figure*}
    \includegraphics[width = 0.75 \textwidth]{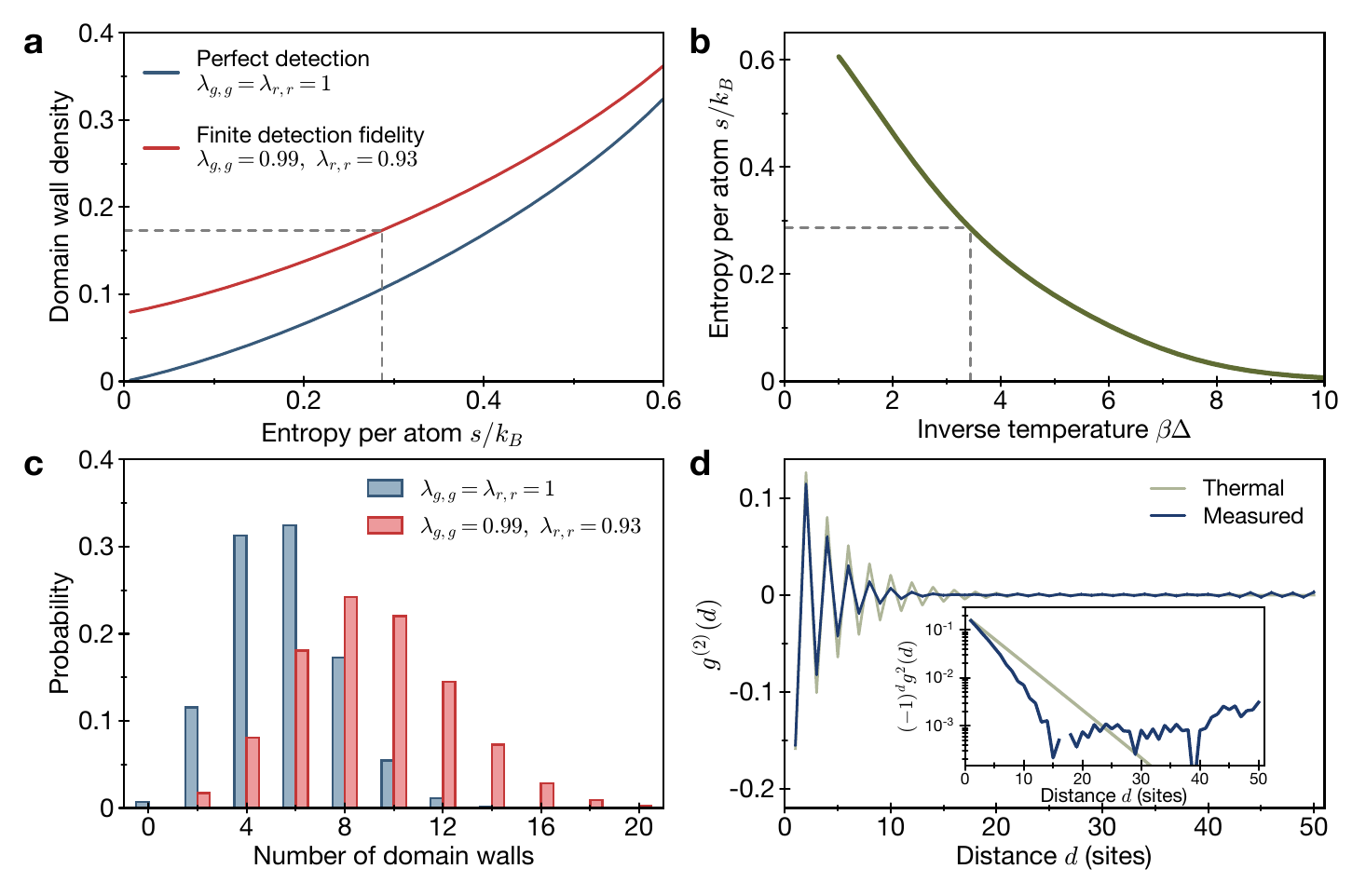}
    \caption{\textbf{Comparison to a thermal state.} \textbf{a,} Domain-wall
    density for thermal states at different entropy per atom $s/k_B$. The lower line
    corresponds to the actual number of domain walls in a system of the
    corresponding temperature; the upper line gives the domain-wall density that would
    be measured at this temperature, given the finite detection fidelity. The
    horizontal dashed line denotes the experimentally measured domain-wall density,
    from which we infer a corresponding entropy per atom and equivalently,
    temperature, in a thermal ensemble. \textbf{b,} Entropy per atoms for a thermal
    state at given inverse temperature $\beta=1/(k_BT)$ in a 51-atom array.
    \textbf{c,} Expected distribution of the number of domain walls for the thermal
    ensemble at $\beta=3.44/\Delta$, with (red) and without (blue) taking into
    account finite detection fidelity. \textbf{d,} Experimentally measured
    correlation function $g^{(2)}(d)$ and correlation function corresponding to a
    thermal ensemble at $\beta=3.44/\Delta$. The inset shows the rectified
    correlation function on a logarithmic scale, indicating that the measured
    correlation function decays exponentially, but with a different correlation
    length than one obtains from a thermal state with the measured number of domain
    walls. \\ \\ \\ \\ \\ \\ \\ \\ \\ \\ \\ \\ \\ \\ \\ \\ \\}
    \label{fig:ED8}
\end{figure*}

\begin{figure*}
    \includegraphics[width = 0.45 \textwidth]{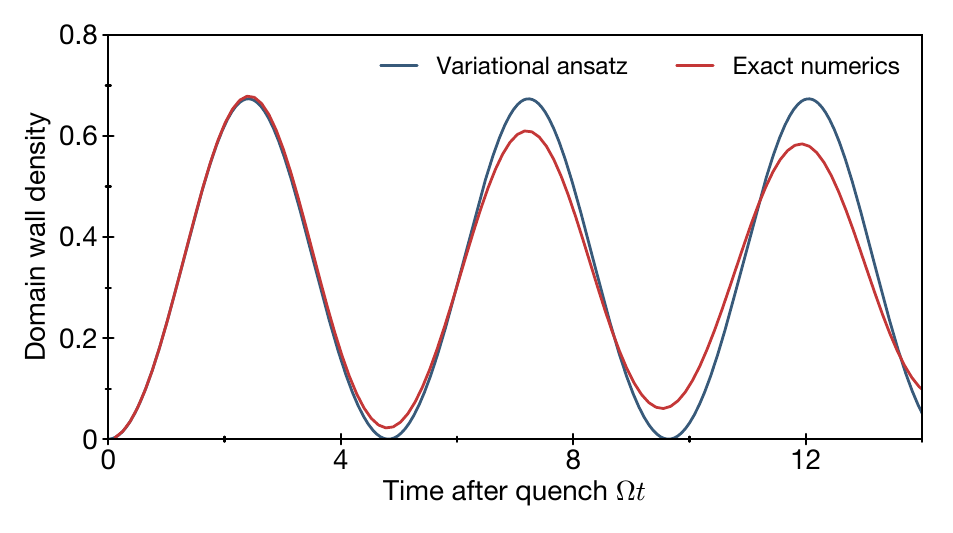}
    \caption{\textbf{Oscillations in domain-wall density: Using a variational
    matrix product state ansatz.} The dynamics of the domain-wall density in the
    bulk of the array under the constrained Hamiltonian $\ham_{\rm c}$ at
    $\Delta=0$ is shown. The blue line shows the evolution of the domain-wall density
    obtained by integrating the variational equation of motion
    (equation~\eqref{MPS_EOM}) with initial conditions $\theta_a=\pi/2$, $\theta_b=0$,
    that is, the crystalline initial state. The red line shows the exact dynamics of
    the domain wall density at the center of a system of 25 atoms initially in the
    crystalline state under the constrained Hamiltonian $\ham_{\rm c}$. \\ \\ \\ \\
    \\ \\ \\ \\ \\ \\ \\ \\ \\ \\ \\ \\ \\}
    \label{fig:ED9}
\end{figure*}

\begin{figure*}
    \includegraphics[width = 0.6 \textwidth]{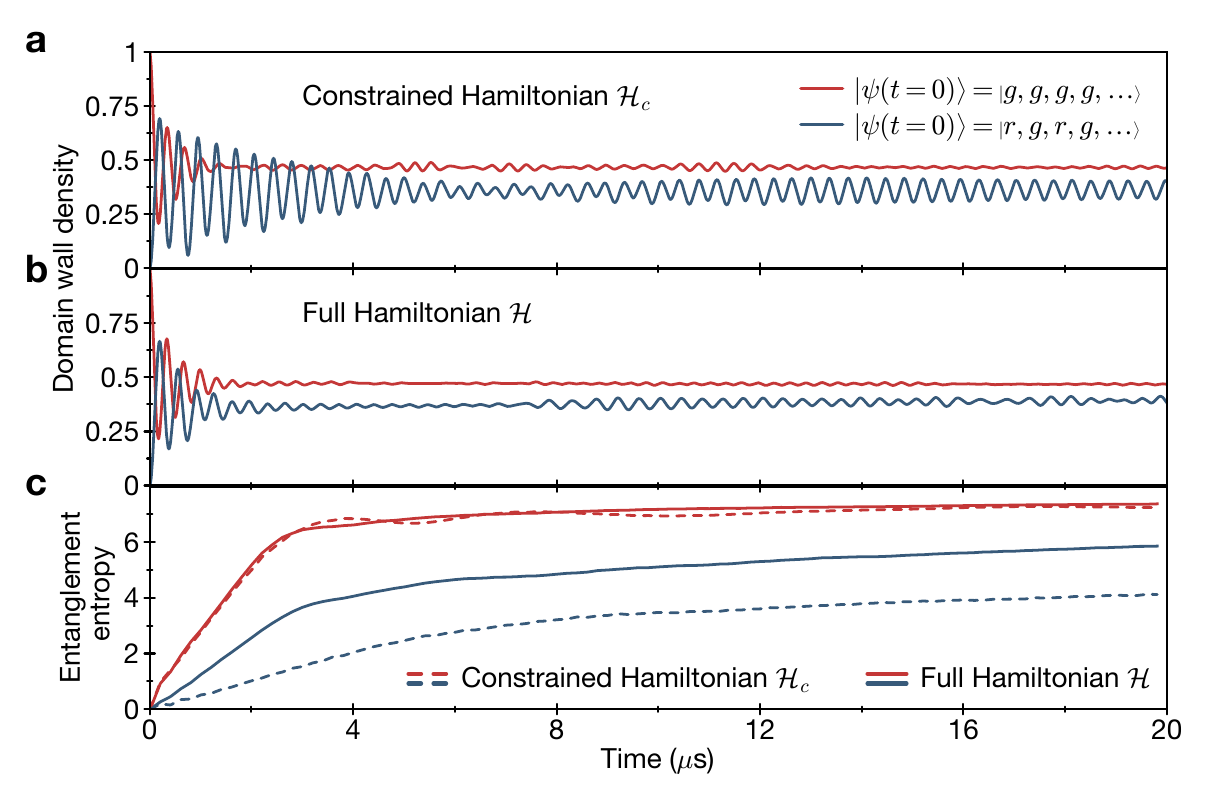}
    \caption{\textbf{Decay of oscillations after a quench and entropy growth.}
    \textbf{a,} Dynamics of the domain-wall density under the constrained
    Hamiltonian $\ham_{\rm c}$ for different initial states. The red line shows the
    domain-wall density for a system of 25 atoms initially prepared in the
    electronic ground state. In this case, the domain-wall density relaxes quickly
    to a steady value corresponding to thermalization. In contrast, the blue line
    shows the dynamics if the system is initialized in the Z$_2$ ordered state. The
    domain-wall density oscillates over several periods and even for very long
    times does not fully relax to a steady value. \textbf{b,} Same as in \textbf{a}
    but taking into account the full $1/R^6$ interactions. While the dynamics for
    an initial state $\ket{g}^{\otimes N}$ is very similar to the one obtained in
    the constrained case, for the crystalline initial state the decay of the
    oscillations is faster than in the constrained model. \textbf{c,} Growth of
    entanglement entropy in a bipartite splitting of the 25-atom array for the
    different cases displayed in \textbf{a} and \textbf{b}. The entropy is defined
    as the von Neumann Entropy of the reduced state of the first 13 atoms of the
    array. The dashed lines correspond to dynamics under the constrained
    Hamiltonian, neglecting the $1/R^6$ tail, whereas the solid lines take the full
    interactions into account. Red lines correspond to the initial state
    $\ket{g}^{\otimes N}$, while blue lines correspond to crystalline initial
    states. In all panels we chose $\Omega=2\pi\times 2\,$ MHz and, where
    applicable, interaction parameters such that the nearest-neighbour interaction
    evaluates to $V_{i,i+1}=2\pi\times 25.6\, \textrm{MHz}$. \\ \\ \\ \\ \\ \\ \\
\\ \\ \\ \\ \\ \\ \\ \\ \\ \\}
	\label{fig:ED10}
\end{figure*}

\end{footnotesize}

\end{document}